\documentclass[amsthm]{elsart}

\usepackage{yjsco}
\usepackage{natbib}

\usepackage{amsfonts}
\usepackage{dsfont}

\usepackage{array,longtable}
\renewcommand*{\arraystretch}{1.5}

\usepackage{amssymb}

\usepackage{amsmath}

\usepackage{amsthm}

\usepackage{graphicx}

\usepackage{epstopdf}

\usepackage[hidelinks]{hyperref}

\usepackage{epsfig}
\usepackage{mdframed}

\usepackage{float}

\usepackage{subcaption}

\usepackage[singlelinecheck=off]{caption}

\usepackage{url}

\PassOptionsToPackage{hyphens}{url}\usepackage{hyperref}

\newcommand{\squeezedown}{\vspace{2.5mm}}

\begin{document}

\begin{frontmatter}

\title{Formal Analysis of Continuous-time Systems using Fourier Transform}

\author{Adnan Rashid}
\address{School of Electrical Engineering and Computer Science (SEECS) \\
National University of Sciences and Technology (NUST), Islamabad, Pakistan}
\ead{adnan.rashid@seecs.nust.edu.pk}

\author{Osman Hasan}
\address{School of Electrical Engineering and Computer Science (SEECS) \\
National University of Sciences and Technology (NUST), Islamabad, Pakistan}
\ead{osman.hasan@seecs.nust.edu.pk}

\begin{abstract}
To study the dynamical behaviour of the engineering and physical systems, we often need to capture their continuous behaviour, which is modeled using differential equations, and perform the frequency-domain analysis of these systems. Traditionally, Fourier transform methods are used to perform this frequency domain analysis using paper-and-pencil based analytical techniques or computer simulations. However, both of these methods are error prone and thus are not suitable for analyzing systems used in safety-critical domains, like medicine and transportation. In order to provide an accurate alternative, we propose to use higher-order-logic theorem proving to conduct the frequency domain analysis of these systems. For this purpose, the paper presents  a higher-order-logic formalization of Fourier transform using the HOL-Light theorem prover. In particular, we use the higher-order-logic based formalizations of differential, integral, transcendental and topological theories of multivariable calculus to formally define Fourier transform and reason about the correctness of its classical properties, such as existence, linearity, time shifting, frequency shifting, modulation, time scaling, time reversal and differentiation in time domain, and its relationships with Fourier Cosine, Fourier Sine and Laplace transforms. We use our proposed formalization for the formal verification of the frequency response of a generic n-order linear system, an audio equalizer and a MEMs accelerometer, using the HOL-Light theorem prover.
\end{abstract}

\begin{keyword}
Frequency Response, Continuous-time Systems, Theorem Proving, Higher-order Logic, HOL-Light, Fourier Transform. \end{keyword}

\end{frontmatter}

\section{Introduction} \label{SEC:Intro}

Fourier Transform~\citep{bracewell1965fourier} is a transform method, which converts a time varying function to its corresponding $\omega$-domain representation, where $\omega$ is its corresponding angular frequency~\citep{beerends2003fourier}. This transformation allows replacing the differentiation and integration in time domain analysis to multiplication and division operators in the frequency domain, which can be easily manipulated. Moreover, the $\omega$-domain representations of the differential equations can also be used for the frequency response analysis of the corresponding systems. Due to these distinguishing features, Fourier transform has been widely used for analyzing many continuous-time systems, such as signal~\citep{papoulis1977signal,gaydecki2004foundations}, and image~\citep{dougherty2009digital} processing algorithms, analog circuits~\citep{thomas2016analysis}, communication systems~\citep{ziemer2006principles,du2010wireless}, medical sciences~\citep{bracewell1965fourier,dougherty2009digital}, mechanical systems~\citep{oppenheim1996signals} and optics~\citep{gaskill1978linear,stark2012application}.

The first step in the Fourier transform based analysis of a continuous-time system,
is to model the dynamics of the system using a differential equation. This differential equation is then transformed to its equivalent $\omega$-domain representation by using the Fourier transform. Next, the resulting $\omega$-domain equation is simplified using various Fourier transform properties, such as existence, linearity, frequency shifting, modulation, time shifting, time scaling, time reversal and differentiation. The main objective of this simplification is to either solve the differential equation to obtain values for the variable $\omega$ or obtain the frequency response of the system corresponding to the given differential equation. The frequency response can in turn be used to analyze the dynamics of the system by studying the impact of different frequency components on the intended behaviour of the given system. The information sought from this analysis plays a vital role in analyzing reliable and performance efficient engineering systems.

Traditionally, the analysis of continuous-time systems, using transform methods, has been done using the paper-and-pencil based analytical technique. However, due to the highly involved human manipulation, the analysis process is error prone, especially when dealing with larger systems, and hence an accurate analysis cannot be guaranteed. Moreover, this kind of manual manipulation does not guarantee that each and every assumption required in the mathematical analysis is written down with the analysis. Thus, some vital assumptions may not accompany the final result of the analysis and a system designed based on such a result may lead to bugs later on.
For example, the Air France Flight 447 crashed in 2009, which resulted in 228 deaths, was attributed to the faulty warning system consisting of speed sensors. These sensors gave wrong/invalid reading about the speed of the airplane, which led to the crash. A more rigourous analysis of the warning system could have prevented this incident.

Computer-based methods including the numerical methods and the symbolic techniques, provide a more scalable option to analyze larger systems. Some of the computer tools involved in these analysis are MATLAB~\citep{MATLAB2016webref}, Mathematica~\citep{wolfram2015mathematica} and Maple~\citep{Maple2016webref}. The numerical analysis involves the approximation of the continuous expressions or the continuous values of the variables due to the finite precision of computer arithmetic, which compromises the accuracy of the analysis. Moreover, it involves a finite number of iterations, depending on the computational resources and computer memory, to judge the values of unknown continuous parameters, which introduces further inaccuracies in the analysis as well.
Similarly, the symbolic tools cannot assure absolute accuracy as they involve discretization of integral to summation while evaluating the improper integral in the definition of Fourier transform~\citep{taqdees2013formalization}.
Moreover, they also contain some unverified symbolic algorithms in their core~\citep{duran2013misfortunes}, which puts another question mark on the accuracy of the results.
Given the widespread usage of the continuous-time systems in many safety-critical domains, such as medicine and transportation, we cannot rely on these above-mentioned analysis methods as the analysis errors could lead to disastrous consequences, including the loss of human lives.

Formal methods~\citep{hasan2015formal} are computer based mathematical techniques that involve the mathematical modeling of the given system and the formal verification of its intended behaviour as a mathematically specified property, which is expressed in an appropriate logic. This verification of the properties of the underlying system is based on mathematical reasoning. Moreover, the mathematical nature of the system model and the desired property guarantees the accuracy of formal analysis. Formal methods have been widely used for the verification of software~\citep{schumann2001automated} and hardware~\citep{camilleri1986hardware} systems and the formalization (or mathematical modeling) of classical mathematics~\citep{hales2005introduction,avigad2014formally}.

Higher-order-logic theorem proving~\citep{harrison2009handbook} is a widely-used formal verification method that has been extensively used to completely analyze continuous systems by leveraging on the high expressiveness of higher-order logic and the soundness of theorem proving. \textit{Umair et al.} formalized the Z-transform~\citep{siddique2014formalization} and used them to analyze an Infinite Impulse Response (IIR) filter. Similarly, \textit{Hira et al.} formalized the Laplace transform ~\citep{taqdees2013formalization} and used their formalization to analyze a Linear Transfer Converter (LTC) circuit.
However, the formalization of Z-transform can only be utilized for the discrete-time system analysis. On the other hand, the formalization of Laplace transform can be used to reason about the solutions of ordinary differential equations and the transfer function analysis of the continuous-time systems~\citep{taqdees2013formalization}, but is only limited to causal functions, i.e., the functions that fulfill the condition: $f(x) = 0$ for all $x < 0$. However, many physical and engineering systems exhibit the non-causal continuous behaviors, involving functions with infinite extent. For example, in optics, the optical image of a point source of light may be described theoretically by a Gaussian function of the form $e^{-x^2}$, which exists for all \emph{x}~\citep{goodman2005introduction}. Another example is the rate of flow of water out of a tap at the bottom of a bucket of water, which can be modeled using $e^{-kt}$, where \emph{t} ranges over the whole real line~\citep{thibos2003fourier}. Fourier transform can cater for the analysis involving both continuous and non-causal functions and thus can overcome the above-mentioned limitations of Z and Laplace transforms.

\begin{figure}[H]
\centering
\scalebox{0.25}
{\includegraphics[trim={0 0.0cm 0 0.0cm},clip]{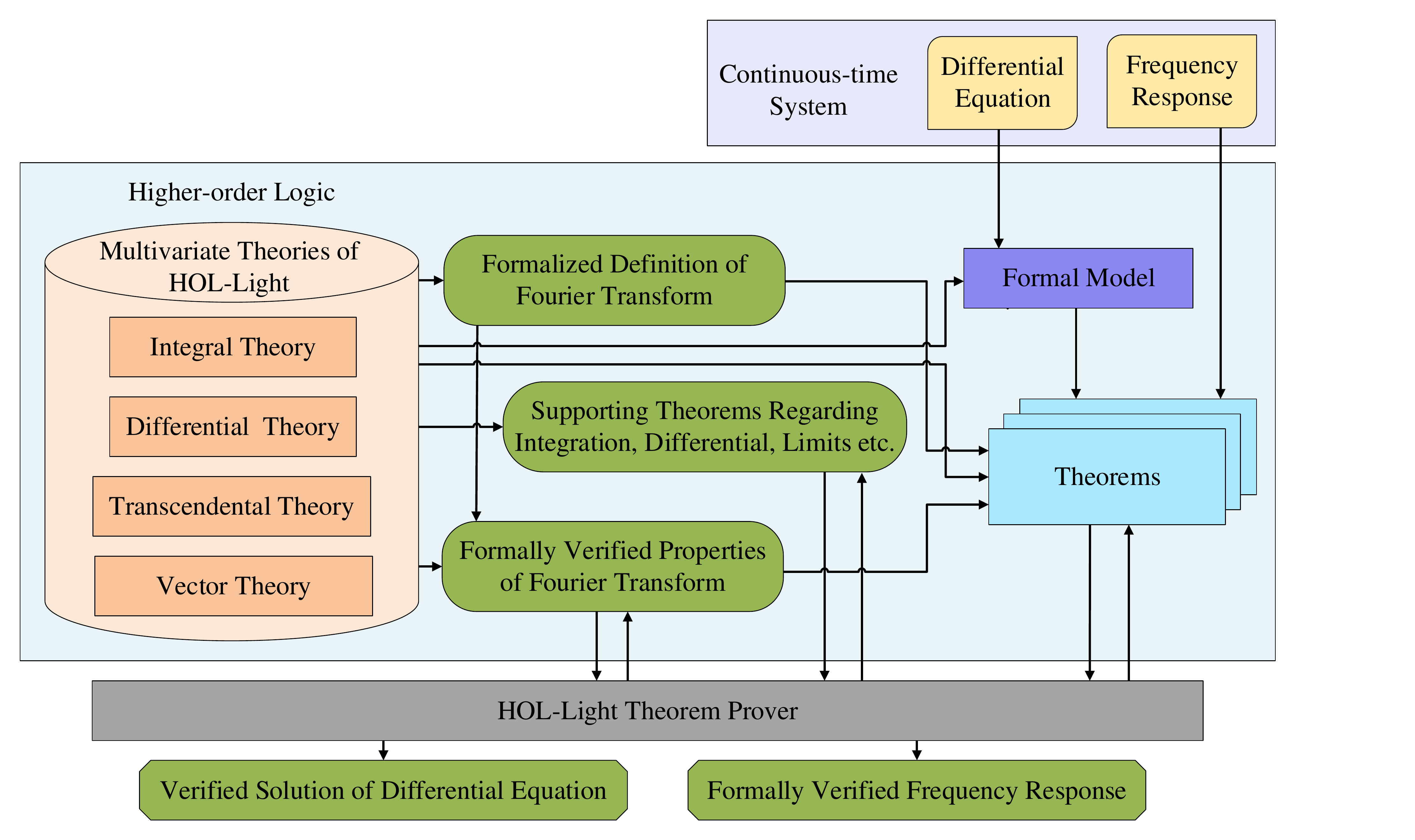}}
\caption{Proposed Framework}
\label{FIG:proposed_methodology}
\end{figure}

With the objective of extending the scope of the analysis based on theorem proving, to cover non-causal functions, we present a higher-order-logic based formalization of Fourier transform in this paper.
In particular, we formalize the definition of Fourier transform in higher-order logic and use it to verify the classical properties of Fourier transform, such as existence, linearity, time shifting, frequency shifting, modulation, time scaling, time reversal, differentiation, and its relations to Fourier Cosine, Fourier Sine and Laplace transforms. These foundations can be built upon to reason about the analytical solutions of differential equations or frequency responses of the physical systems, as depicted in Fig.~\ref{FIG:proposed_methodology}. The user of the proposed formal analysis framework is required to develop a formal model of the given system, by using its corresponding differential equation. Similarly, the desired frequency response behavior from the given system can be captured as a proof goal (theorem) that can be formed using this behavior along with the formal model of the given system. The proof goal can then be verified based on the above-mentioned formalization of Fourier transform within the sound core of the HOL-Light theorem prover, which is an interactive theorem proving environment for conducting proofs in higher-order logic. The availability of the above-mentioned formally verified properties decreases the manual user interaction and thus effort required while performing the formal Fourier transform based analysis of a system. Besides the above-mentioned foundational formalization of Fourier transform, we also use these results to verify a relationship of frequency response of a generic n-order system~\citep{adams2012continuous}. This generic relationship can be specialized to facilitate the reasoning process of the formal frequency response analysis of any specific order system.
In order to illustrate the practical utilization of the proposed formalization, we present a formal analysis of an audio equalizer~\citep{adams2012continuous} and a MEMs accelerometer~\citep{kaajakari2009practical},
which are extensively used in communication systems and many safety critical systems, respectively.
We use the HOL-Light theorem prover~\citep{harrison-hol-light} for the proposed formalization in order to build upon its comprehensive reasoning support for multivariable calculus. Particularly, the proposed formalization heavily relies upon the formalization of differential, integration, topological and transcendental theories of multivariable calculus.

The rest of the paper is organized as follows: Section \ref{SEC:Preliminaries} provides a brief introduction about the HOL-Light theorem prover and the multivariable calculus theories of HOL-Light.
%We present the proposed framework for the formal analysis of the continuous-time systems in Section \ref{SEC:proposed_methodology}.
Section \ref{SEC:Formalization_of_Fourier} presents the formalization of the Fourier transform definition and the conditions required for its existence. We provide the verification of the classical properties of Fourier transform in Section \ref{SEC:Formal_verif_Fourier_properties}. Section \ref{SEC:fourier_trans_comm_used_funs} presents the Fourier transforms of some commonly used functions. We present the formal analysis of generic n-order system in Section~\ref{SEC:formal_analysis_generic_n_order_sys}.
Section \ref{SEC:applications} provides the verification of the frequency response of
an audio equalizer and a MEMs accelerometer.
Section~\ref{SEC:discussion} presents the discussion, which highlights upon the main challenges faced in the proposed formalization.
Finally, Section \ref{SEC:Conclusion} concludes the paper.

\section{Preliminaries} \label{SEC:Preliminaries}

In this section, we present an introduction to the HOL-Light theorem prover and an overview about the multivariable calculus theories of HOL-Light, which provide the foundational support for the proposed formalization.

\subsection{HOL-Light Theorem Prover} \label{SUBSEC:HOL_Light_theorem_prover}

HOL-Light~\citep{harrison-hol-light} is an interactive theorem proving environment for conducting proofs in higher-order logic. The logic in the HOL-Light system is represented in the strongly-typed functional programming language ML~\citep{paulson_96}. A theorem is a formalized statement that may be an  axiom or could be deduced from already verified  theorems by an inference rule. A theorem  consists of a finite set $\Omega$ of Boolean terms, called the assumptions, and  a Boolean term $S$, called the conclusion. Soundness is assured as every new theorem must be verified by applying the basic axioms and primitive inference rules or any other previously verified theorems/inference rules. A HOL-Light theory is a collection of valid HOL-Light types, constants, axioms, definitions and theorems.
Various mathematical foundational concepts have been formalized and saved as HOL-Light theories.
The HOL-Light theorem prover provides an extensive support of theorems regarding, Boolean algebra, arithmetic, real numbers, transcendental functions and multivariate analysis such as differential, integration, vectors and topology, in the form of theories, which are extensively used in our formalization. In fact, one of the primary reasons to chose the HOL-Light theorem prover for the proposed formalization was the presence of an extensive support of multivariable calculus theories.
There are many automatic proof procedures and proof assistants~\citep{Harrison_formalized_mathematics} available in HOL-Light, which help the user in concluding a proof more efficiently.

Table~\ref{TAB:Hol_light_symbols} presents the standard and HOL-Light representations and the meanings of some commonly used symbols in this paper.

\begin{table}[h]
\flushleft
\caption{HOL-Light Symbols}
\label{TAB:Hol_light_symbols}
  \resizebox{1.0\textwidth}{!}{\begin{minipage}{\textwidth}
{\renewcommand{\arraystretch}{1.005}% for the vertical padding
\begin{tabular}{p{3.05cm} p{4.1cm} p{5.1cm}}
\hline\hline
 HOL-Light Symbols & Standard Symbols  & Meanings  \\ \hline \hline
 $\mathtt{/\backslash}$ & and & Logical $and$ \\ \hline
    $\mathtt{\backslash/}$ & or & Logical $or$  \\ \hline
 $\mathtt{\sim}$ & not & Logical $negation$ \\ \hline
%    $\mathtt{T}$ & true & Logical true value  \\ \hline
% $\mathtt{F}$ & false & Logical false value  \\ \hline
    $\mathtt{==>}$ & $ \longrightarrow$ &  Implication\\ \hline
 $\mathtt{<=>}$ & $ = $ &  Equality in Boolean domain\\ \hline
    $\mathtt{!x. t}$ & $ \forall x.t$ &  For all $x$ : $t$  \\ \hline
 $\mathtt{?x. t}$ & $ \exists x.t$ &  There exists $x$ : $t$  \\ \hline
 \texttt{$\lambda$x.t} & $\lambda x.t$ & Function that maps $x$ to $t(x)$  \\ \hline
    $\mathtt{num}$  &  $\{0,1,2,\ldots\}$ & Positive Integers data type   \\ \hline
 $\mathtt{real}$ &  All Real numbers &  Real data type  \\ \hline
    $\mathtt{SUC\ n}$&  ($n + 1$)& Successor of natural number \\ \hline
 $\mathtt{\& a}$  & $\mathbb{N} \rightarrow \mathbb{R}$ & Typecasting  from Integers to Reals    \\ \hline
    $\mathtt{abs\ x}$ & $|x|$  & Absolute function   \\ \hline
% $\mathtt{HD\ l}$  &  $head$ & Head element of list $l$ \\ \hline
%    $\mathtt{TL\ l}$  &  $tail$ & Tail of list $l$\\ \hline
 $\mathtt{EL\ n\ l}$  &  $element$ & $n^{th}$ element of list l \\ \hline
%    $\mathtt{LENGTH\ l}$  &  $length$ & Length of list $l$\\ \hline
%    $\mathtt{Cx (a)}$  & $\mathbb{R} \rightarrow \mathbb{C}$ & Typecasting  from Reals to Complex    \\ \hline
\end{tabular}
  }
%   }
      \end{minipage}}
\end{table}

\subsection{Multivariable Calculus Theories in HOL-Light} \label{SEC:Mult_cal_theories}

A N-dimensional vector is represented as a $\mathds{R^N}$ column matrix with each of its element as a real number in HOL-Light~\citep{harrison2013hol}. All of the vector operations are thus performed using matrix manipulations. A complex number is defined as a 2-dimensional vector, i.e., a $\mathds{R}^2$ column matrix. All of the multivariable calculus theorems are verified in HOL-Light for functions with an arbitrary data-type $\mathds{R^N} \rightarrow \mathds{R^M}$.

Some of the frequently used HOL-Light functions in our work are explained below:

\begin{defn}
\label{DEF:cx_and_ii}
\emph{Cx and ii} \\{\small
\textup{\texttt{$\vdash$ $\forall$ a. Cx a = complex (a, \&0) \\
$\mathtt{}$$\vdash$ ii = complex (\&0, \&1)
}}}
\end{defn}

\noindent $\mathtt{Cx}$ is a type casting function from real ($\mathds{R}$) to complex ($\mathds{R}^2$). It accepts a real number and returns its corresponding complex number with the imaginary part equal to zero, where the $\texttt{\&}$ operator type casts a natural number ($\mathds{N}$) to its corresponding real number ($\mathds{R}$). Similarly, $\mathtt{ii}$ (iota) represents a complex number having the real part equal to zero and the magnitude of the imaginary part equal to 1.

\begin{defn}
\label{DEF:re_im_lift_drop}
\emph{Re, Im, lift and drop} \\{\small
\textup{\texttt{$\vdash$ $\forall$ z. Re z = z\$1 \\
$\mathtt{}$$\vdash$ $\forall$ z. Im z = z\$2 \\
$\mathtt{}$$\vdash$ $\forall$ x. lift x = (lambda i. x) \\
$\mathtt{}$$\vdash$ $\forall$ x. drop x = x\$1
}}}
\end{defn}

The function $\mathtt{Re}$ accepts a complex number and returns its real part. Here, the notation $\mathtt{z\$i}$ represents the $i^{th}$ component of vector $\texttt{z}$. Similarly, $\mathtt{Im}$ takes a complex number and returns its imaginary part. The function $\mathtt{lift}$ accepts a variable of type $\mathds{R}$ and maps it to a 1-dimensional vector with the input variable as its single component. It uses the \texttt{lambda} operator in HOL to construct a vector based on its components~\citep{harrison2013hol}. Similarly, $\mathtt{drop}$ takes a 1-dimensional vector and returns its single element as a real number. In order to make the functions $\mathtt{lift}$ and $\mathtt{drop}$ better understandable to a non-HOL user, we use $\mathtt{\overline{x}}$ and $\mathtt{\underline{x}}$ as the equivalent symbols for \texttt{lift x} and \texttt{drop x}, respectively.

\begin{defn}
\label{DEF:exp_ccos_csine}
\emph{Exponential, Complex Cosine and Sine Functions} \\{\small
\textup{\texttt{$\vdash$ $\forall$ x. exp x = Re (cexp (Cx x)) \\
$\mathtt{}$$\vdash$ $\forall$ z. ccos z = (cexp (ii $\ast$ z) + cexp (--ii $\ast$ z)) / Cx (\&2)   \\
$\mathtt{}$$\vdash$ $\forall$ z. csin z = (cexp (ii $\ast$ z) - cexp (--ii $\ast$ z)) / (Cx (\&2) $\ast$ ii)
}}}
\end{defn}

The complex exponential and real exponentials are represented as $\texttt{cexp}:\mathds{R}^2 \rightarrow \mathds{R}^2$
and $\mathtt{exp}:\mathds{R} \rightarrow \mathds{R}$ in HOL-Light, respectively. Similarly, the complex cosine $\mathtt{ccos}$ and complex sine $\mathtt{csin}$ functions are formally defined in terms of $\texttt{cexp}$ using the Euler's formula~\citep{hol_light2016transcendentals}.

\begin{defn}
\label{DEF:vector_integral}
\emph{Vector Integral and Real Integral} \\
{\small
\textup{\texttt{$\vdash$ $\forall$ f i. integral i f = (@y. (f has\_integral y) i)
}}} \\
{\small
\textup{\texttt{$\vdash$ $\forall$ f i. real\_integral i f = (@y. (f has\_real\_integral y) i)
}}}
\end{defn}

The function $\mathtt{integral}$ represents the vector integral and is defined using the Hilbert choice operator $\texttt{@}$ in the functional form. It takes the integrand function $\texttt{f}$, having an arbitrary type $\mathds{R}^N \rightarrow \mathds{R}^M$, and a vector-space $\mathtt{i}: \mathds{R}^N \rightarrow \mathds{B}$, which defines the region of convergence as $\mathds{B}$ represents the boolean data type, and returns a vector $\mathds{R}^M$ which is the integral of $\mathtt{f}$ on $\mathtt{i}$. The function $\mathtt{has\_integral}$ represents the same relationship in the relational form.
It is a predicate, which accepts the integrand function, integral value and the region of integration, and returns \texttt{T} if the integral of the integrand function over the region of integration is equal to the integral value. Where as, the function $\mathtt{integral}$ accepts the integrand function, its region of integration and returns the value of the integral over the given region of the integration using the Hilbert choice operator $\texttt{@}$.
Similarly, the function $\mathtt{real\_integral}$ accepts the integrand function $\mathtt{f} : \mathds{R} \rightarrow \mathds{R}$ and a set of real numbers $\mathtt{i}: \mathds{R} \rightarrow \mathds{B}$ and returns the real-valued integral of the function $\mathtt{f}$ over $\mathtt{i}$.
The region of integration, for both of the above integrals can be defined to be bounded by a vector interval $[a, b]$ or real interval $[a, b]$ using the HOL-Light functions $\mathtt{interval \ [a,b]}$ and $\mathtt{real\_interval \ [a,b]}$, respectively.

\begin{defn}
\label{DEF:vector_derivative}
\emph{Vector Derivative and Real Derivative} \\
{\small
\textup{\texttt{$\vdash$ $\forall$ f net. vector\_derivative f net = (@f'. (f has\_vector\_derivative f') net)
}}} \\
{\small
\textup{\texttt{$\vdash$ $\forall$ f x. real\_derivative f x = (@f'. (f has\_real\_derivative f') (atreal x))
}}}

\end{defn}

The function $\mathtt{vector\_derivative}$ takes a function $\texttt{f} : \mathds{R}^1 \rightarrow \mathds{R}^M$ and a $\texttt{net} : \mathds{R}^1 \rightarrow \mathds{B}$, which defines the point at which $\texttt{f}$ has to be differentiated, and returns a vector of data-type $\mathds{R}^M$, which represents the differential of $\texttt{f}$ at $\texttt{net}$.
Moreover, depending on the usage of the definition, \texttt{net} can be specified as (\texttt{at a within s}) or (\texttt{at a}), which can be a point $\texttt{a} : \mathds{R}^1$ of a set $\texttt{s} : \mathds{R}^1 \rightarrow \mathds{B}$ or point $\texttt{a} : \mathds{R}^1$, respectively, where the function \texttt{f} has to be differentiated.
The function $\mathtt{has\_vector\_derivative}$ defines the same relationship in the relational form.
Similarly, the function $\mathtt{real\_derivative}$ accepts a function $\texttt{f} : \mathds{R} \rightarrow \mathds{R}$ and a real number $\texttt{x}$, which is the point at which $\texttt{f}$ has to be differentiated, and returns a variable of data-type $\mathds{R}$, which represents the differential of $\texttt{f}$ at $\texttt{x}$. The function $\mathtt{has\_real\_derivative}$ defines the same relationship in the relational form.

\begin{defn}
\label{DEF:limit_of_function}
\emph{Limit of a function} \\{\small
\textup{\texttt{$\vdash$ $\forall$ f net. lim net f = (@l. (f $\rightarrow$ l) net)
}}}
\end{defn}

The function $\mathtt{lim}$ accepts a $\texttt{net}$ with elements of arbitrary data-type $\mathds{A}$ and a function $\texttt{f} : \mathds{A} \rightarrow \mathds{R}^M$ and returns $\texttt{l}$ of data-type $\mathds{R}^M$, i.e., the value to which $\texttt{f}$ converges at the given $\texttt{net}$. Moreover, \texttt{net} can take \texttt{at\_posinfinity} or \texttt{at\_neginfinity} to model the positive infinity or negative infinity, respectively.

We build upon the above-mentioned fundamental functions of multivariable calculus to formalize the Fourier transform in the next section.

\section{Formalization of Fourier Transform} \label{SEC:Formalization_of_Fourier}

The Fourier transform of a function $f(t)$ is mathematically defined as:

\begin{equation}\label{EQ:fourier_transform}
\mathcal{F} [f (t)] = F(\omega) = \int_{-\infty}^{+\infty} {f(t)e^{-i \omega t}} dt, \ \omega \ \epsilon \  \mathds{R}
\end{equation}

\squeezedown

\noindent where $f$ is a function from $\mathds{R}^1 \rightarrow \mathds{C}$ and $\omega$ is a real variable. The limit of integration is from ${-\infty}$ to ${+\infty}$. We formalize Equation~\ref{EQ:fourier_transform} in HOL-Light as follows:

\begin{defn}
\label{DEF:fourier_transform}
\emph{Fourier Transform} \\{\small
\textup{\texttt{$\vdash$ $\forall$ w f. fourier\_transform f w = \\
$\mathtt{\ }$\hspace{2.80cm} integral UNIV ($\lambda$t. cexp (--((ii $\ast$ Cx w) $\ast$ Cx $\mathtt{\underline{t}}$)) $\ast$ f t)
}}}
\end{defn}

The function \texttt{fourier\_transform} accepts a complex-valued function $ \texttt{f}: \mathds{R}^1 \rightarrow \mathds{R}^2 $ and a real number $\texttt{w}$ and returns a complex number that is the Fourier transform of $ \texttt{f} $ as represented by Equation \ref{EQ:fourier_transform}. In the above function, we used complex exponential function $ \texttt{cexp}: \mathds{R}^2 \rightarrow \mathds{R}^2 $ because the return data-type of the function $ \texttt{f} $ is $ \mathds{R}^2 $. To multiply $\texttt{w}$ with $ \texttt{ii} $, we first converted $\texttt{w}$ into a complex number $ \mathds{R}^2 $ using $ \texttt{Cx} $. Similarly, the data-type of $ \texttt{t} $ is $ \mathds{R}^1 $ and to multiply it with $ \mathtt{ii \ast Cx \ w} $, it is first converted into a real number $\mathtt{\underline{t}}$ by using $\texttt{drop}$ and then it is converted to data-type $ \mathds{R}^2 $ using $ \texttt{Cx} $. Next, we use the vector function $ \texttt{integral} $ (Definition \ref{DEF:vector_integral}) to integrate the expression $ f(t)e^{-i \omega t} $ over the whole real line since the data-type of this expression is $ \mathds{R}^2 $. Since the region of integration of the vector integral function must be a vector space therefore we represented the interval of the integral by $ \texttt{UNIV}:\mathds{R}^1 $, which represents the whole real line.

The Fourier transform of a function $ f $ exists, i.e., the integrand of Equation \ref{EQ:fourier_transform} is integrable, and the integral has some converging limit value, if $f$ is piecewise smooth and is absolutely integrable on the whole real line~\citep{rashid2016formalization,beerends2003fourier,rashid2017tmformalization}. A function is said to be piecewise smooth on an interval if it is piecewise differentiable on that interval.
The Fourier existence condition can thus be formalized in HOL-Light as follows:

\begin{defn}
\label{DEF:fourier_exists}
\emph{Fourier Exists} \\{\small
\textup{\texttt{$\vdash$ $\forall$ f. fourier\_exists f $\Leftrightarrow$ \\
$\mathtt{\ }$\hspace{0.5cm} ($\forall$ a b. f piecewise\_differentiable\_on interval [$\mathtt{\overline{a}}$, $\mathtt{\overline{b}}$]) $\wedge$ \\
$\mathtt{\ }$\hspace{1.85cm} f absolutely\_integrable\_on UNIV
}}}
\end{defn}

\noindent In the above function, the first conjunct expresses the piecewise smoothness condition for the function $\texttt{f}$.
Whereas, the second conjunct represents the condition that the function $\texttt{f}$ is absolutely integrable on the whole real line.
Next, we present a very important property of the Fourier existence as follows:

\begin{thm}
\label{THM:linearity_prop_four_exist}
\emph{Linearity of Fourier Existence} \\{\small
\textup{\texttt{$\vdash$ $\forall$ f g a b. fourier\_exists f $\wedge$ fourier\_exists g $\Rightarrow$ \\
$\mathtt{\ }$\hspace{5.0cm} fourier\_exists ($\lambda$x. a $\ast$ f x + b $\ast$ g x)
}}}
\end{thm}

\noindent where $ \texttt{a}: \mathds{C} $ and $ \texttt{b}: \mathds{C} $ are arbitrary constants acting as the scaling factors.
The proof of above theorem is based on the linearity properties of integration, limit and piecewise differentiability.

\vspace*{-2mm}

\section{Formal Verification of Fourier Transform Properties} \label{SEC:Formal_verif_Fourier_properties}

In this section, we use Definitions~\ref{DEF:fourier_transform} and~\ref{DEF:fourier_exists} and Theorem~\ref{THM:linearity_prop_four_exist} to verify some of the classical properties of Fourier transform and its relationship with various transforms, like Fourier Cosine, Fourier Sine, and Laplace transforms and Fourier transform of a $n^{th}$-order differential equation in HOL-Light. The verification of these properties and the relationships not only ensures the correctness of our definitions but also plays a vital role in minimizing the user intervention and time consumption in reasoning about Fourier transform based frequency domain analysis of continuous-time systems, as will be depicted in Section \ref{SEC:applications} of this paper.

\subsection{Properties of Fourier Transform}

The existence of the improper integral of Fourier Transform is a pre-condition for most of the arithmetic manipulations involving the Fourier transform. This condition is formalized in HOL-Light as the following theorem:

%\squeezedown

%\begin{mdframed}
%\begin{flushleft}
\begin{thm}
\label{THM:prop_01_integrable_univ}
\emph{Integrability of Integrand of Fourier Transform Integral} \\{\small
\textup{\texttt{$\vdash$ $\forall$ f w. fourier\_exists f $\Rightarrow$ \\
$\mathtt{\ }$\hspace{1.9cm} ($\lambda$t. cexp (--((ii $\ast$ Cx w) $\ast$ Cx $\mathtt{\underline{t}}$)) $\ast$ f t) integrable\_on UNIV
}}}
\end{thm}
%\end{flushleft}
%\end{mdframed}

%\squeezedown

The proof of Theorem~\ref{THM:prop_01_integrable_univ} is based on splitting the region of integration, i.e., the whole real line $\mathtt{UNIV}:\mathds{R}^1$, as a union of positive real line (interval $[0,\infty)$) and negative real line (interval $(-\infty, 0]$). Next, we split the complex-valued integrand, $f(t)e^{-i \omega t}$, into its corresponding real and imaginary parts. In this process, we need the integrability of the integrand, which can be derived by the piecewise differentiability of \texttt{fourier\_exists} condition.
Finally, some theorems regarding integration, integrability, continuity and some properties of the transcendental functions are used to conclude the proof of Theorem~\ref{THM:prop_01_integrable_univ}.

Next, we verified some of the classical properties of Fourier transform, given in Table~\ref{TAB:properties_of_Fourier_transform}.

%-----------------------------------------------------------------------------------------------------------------------------------------------------------------------
%-----------------------------------------------------------------------------------------------------------------------------------------------------------------------

\begin{scriptsize}
    \begin{longtable}{|p{2cm}|p{3cm}|p{7cm}|p{3cm}|}
\caption{Properties of Fourier Transform}
\label{TAB:properties_of_Fourier_transform}
\endfirsthead
\endhead
    \hline
    \hline
    \multicolumn{1}{l}{Mathematical Form}   &
    \multicolumn{1}{l}{\hspace{-0.4cm} Formalized Form}

     \\ \hline \hline

%%%%%%%%%%%%%%%%%%%%%%%%%%%%%%%%%%%%%%%%%%%

    % Line 1

\multicolumn{2}{c}{\textbf{Linearity}} \\ \hline

   \multicolumn{1}{l}{ {$\begin{array} {lcl} \hspace{-0.2cm} \textit{$ \mathcal{F} [ \alpha f(t) + \beta g(t)] = $ } \\
\textit{$\mathtt{\ }$\hspace{0.4cm} $\alpha F(\omega) + \beta G(\omega) $     }
 \end{array}$}  }  &

   \multicolumn{1}{l}{{ $\begin{array} {lcl} \textup{\texttt{\hspace{-0.4cm}$\vdash$ $\forall$ f g w a b. fourier\_exists f $\wedge$ fourier\_exists g $\Rightarrow$     }} \\
\textup{\texttt{$\mathtt{\ }$\hspace{0.0cm} fourier\_transform ($\lambda$t. a $\ast$ f t + b $\ast$ g t) w =  }} \\
\textup{\texttt{$\mathtt{\ }$\hspace{0.5cm} a $\ast$ fourier\_transform f w + b $\ast$ fourier\_transform g w  }}
 \end{array}$}}    \\ \hline

%%%%%%%%%%%%%%%%%%%%%%%%%%%%%%%%%%%%%%%%%%%

    % Line 2

\multicolumn{2}{c}{\textbf{Time Shifting (Time Advance and Time Delay)}} \\ \hline

    \multicolumn{1}{l}{ {$\begin{array} {lcl} \hspace{-0.2cm} \textit{$ \mathcal{F} [f(t + t0)] = F(\omega)e^{+ i \omega t0} $     }
 \end{array}$} }   &

   \multicolumn{1}{l}{ {$\begin{array} {lcl} \textup{\texttt{\hspace{-0.3cm}$\vdash$ $\forall$ f w t0. fourier\_exists f $\Rightarrow$     }} \\
\textup{\texttt{$\mathtt{\ }$\hspace{-0.2cm} fourier\_transform ($\lambda$t. f (t + t0)) w =  }} \\
\textup{\texttt{$\mathtt{\ }$\hspace{0.0cm} fourier\_transform f w $\ast$ cexp ((ii $\ast$ Cx w) $\ast$ Cx $\mathtt{\underline{t0}}$) \hspace{-1.0cm}  }}
 \end{array}$} }    \\ \hline

%%%%%%%%%%%%

    \multicolumn{1}{l}{  {$\begin{array} {lcl} \hspace{-0.2cm} \textit{$ \mathcal{F} [f(t - t0)] = F(\omega)e^{-i \omega t0} $     }
 \end{array}$} }   &

   \multicolumn{1}{l}{ {$\begin{array} {lcl} \textup{\texttt{\hspace{-0.3cm}$\vdash$ $\forall$ f w t0. fourier\_exists f $\Rightarrow$     }} \\
\textup{\texttt{$\mathtt{\ }$\hspace{-0.2cm} fourier\_transform ($\lambda$t. f (t - t0)) w =  }} \\
\textup{\texttt{$\mathtt{\ }$\hspace{0.0cm} fourier\_transform f w $\ast$ cexp (--((ii $\ast$ Cx w) $\ast$ Cx $\mathtt{\underline{t0}}$)) \hspace{-1.0cm}  }}
 \end{array}$} }    \\ \hline

%%%%%%%%%%%%%%%%%%%%%%%%%%%%%%%%%%%%%%%%%%%

%    % Line 3

\multicolumn{2}{c}{\textbf{Frequency Shifting (Right and Left Shifting)}} \\ \hline

    \multicolumn{1}{l}{ {$\begin{array} {lcl} \hspace{-0.2cm} \textit{$ \mathcal{F} [ e^{+i \omega _0 t} f(t)] = F(\omega - \omega _0) $     }
 \end{array}$} }   &

   \multicolumn{1}{l}{ {$\begin{array} {lcl} \textup{\texttt{\hspace{-0.3cm}$\vdash$ $\forall$ f w w0. fourier\_exists f $\Rightarrow$     }} \\
\textup{\texttt{$\mathtt{\ }$\hspace{-0.2cm} fourier\_transform ($\lambda$t. cexp ((ii $\ast$ Cx w0) $\ast$ Cx $\mathtt{\underline{t}}$) $\ast$ f t) w = \hspace{-1.0cm} }} \\
\textup{\texttt{$\mathtt{\ }$\hspace{0.0cm} fourier\_transform f (w - w0)  }}
 \end{array}$} }    \\ \hline

%%%%%%%%%%%%%

    \multicolumn{1}{l}{  {$\begin{array} {lcl} \hspace{-0.2cm} \textit{$ \mathcal{F} [ e^{-i \omega _0 t} f(t)] = F(\omega + \omega _0) $     }
 \end{array}$} }   &

    \multicolumn{1}{l}{ {$\begin{array} {lcl} \textup{\texttt{\hspace{-0.3cm}$\vdash$ $\forall$ f w w0. fourier\_exists f $\Rightarrow$     }} \\
\textup{\texttt{$\mathtt{\ }$\hspace{-0.4cm} fourier ($\lambda$t. cexp (--(ii $\ast$ Cx w0) $\ast$ Cx $\mathtt{\underline{t}}$) $\ast$ f t) w = \hspace{-1.0cm} }} \\
\textup{\texttt{$\mathtt{\ }$\hspace{0.0cm} fourier f (w + w0)  }}
 \end{array}$}  }    \\ \hline

%%%%%%%%%%%%%%%%%%%%%%%%%%%%%%%%%%%%%%%%%%%

    %Line 4

\multicolumn{2}{c}{\textbf{Modulation (Cosine and Sine Based Modulation)}} \\ \hline

    \multicolumn{1}{l}{    {$\begin{array} {lcl} \hspace{-0.2cm} \textit{$ \mathcal{F}[cos(\omega_0 t) f(t)] = $ } \\
\textit{$\mathtt{\ }$\hspace{-0.2cm} $\dfrac{F(\omega - \omega _0) + F(\omega + \omega _0)}{2} $     }
 \end{array}$}  }   &

    \multicolumn{1}{l}{{$\begin{array} {lcl} \textup{\texttt{\hspace{-0.3cm}$\vdash$ $\forall$ f w w0. fourier\_exists f $\Rightarrow$     }} \\
\textup{\texttt{$\mathtt{\ }$\hspace{-0.4cm} fourier\_transform ($\lambda$t. ccos (Cx w0 $\ast$ Cx $\mathtt{\underline{t}}$) $\ast$ f t) w =  }} \\
\textup{\texttt{$\mathtt{\ }$\hspace{0.0cm} (fourier\_transform f (w - w0) +   }}  \\
\textup{\texttt{$\mathtt{\ }$\hspace{1.5cm} fourier\_transform f (w + w0)) / Cx (\&2)  }}
 \end{array}$}}    \\ \hline

 %%%%%%%%%%%%%%%%

    \multicolumn{1}{l}{    {$\begin{array} {lcl} \hspace{-0.2cm} \textit{$ \mathcal{F}[sin(\omega_0 t) f(t)] = $ } \\
\textit{$\mathtt{\ }$\hspace{-0.2cm} $ \dfrac{F(\omega - \omega _0) - F(\omega + \omega _0)}{2i} $     }
 \end{array}$}  }   &

    \multicolumn{1}{l}{{$\begin{array} {lcl} \textup{\texttt{\hspace{-0.3cm}$\vdash$ $\forall$ f w w0. fourier\_exists f $\Rightarrow$     }} \\
\textup{\texttt{$\mathtt{\ }$\hspace{-0.4cm} fourier\_transform ($\lambda$t. csin (Cx w0 $\ast$ Cx $\mathtt{\underline{t}}$) $\ast$ f t) w =  }} \\
\textup{\texttt{$\mathtt{\ }$\hspace{0.0cm} (fourier\_transform f (w - w0) -  }}  \\
\textup{\texttt{$\mathtt{\ }$\hspace{1.5cm} fourier\_transform f (w + w0)) / (Cx (\&2) $\ast$ ii)  }}
 \end{array}$}}    \\ \hline

%%%%%%%%%%%%%%%%%%%%%%%%%%%%%%%%%%%%%%%%%%%

    %Line 5

\multicolumn{2}{c}{\textbf{Time Scaling}} \\ \hline

    \multicolumn{1}{l}{    $ \mathcal{F}[f(at)] = \dfrac{1}{|a|}F(\dfrac{\omega}{a}) $      }   &

    \multicolumn{1}{l}{{$\begin{array} {lcl} \textup{\texttt{\hspace{-0.3cm}$\vdash$ $\forall$ f w a. fourier\_exists f $\wedge$ $\sim$(a = \&0)  $\Rightarrow$  }} \\
    \textup{\texttt{$\mathtt{\ }$\hspace{0.2cm} fourier\_transform ($\lambda$t. f (a \% t)) w =  }} \\
\textup{\texttt{$\mathtt{\ }$\hspace{0.2cm}  (Cx (\&1) / Cx (abs a)) $\ast$ fourier\_transform f (w / a)  }}
 \end{array}$}}    \\ \hline

%%%%%%%%%%%%%%%%%%%%%%%%%%%%%%%%%%%%%%%%%%%

    %Line 6

\multicolumn{2}{c}{\textbf{Time Reversal}} \\ \hline

    \multicolumn{1}{l}{    $ \mathcal{F}[f(-t)] = F(-\omega) $      }   &

    \multicolumn{1}{l}{{$\begin{array} {lcl} \textup{\texttt{\hspace{-0.3cm}$\vdash$ $\forall$ f w. fourier\_exists f $\Rightarrow$  }} \\
\textup{\texttt{$\mathtt{\ }$\hspace{-0.4cm}  fourier\_transform ($\lambda$t. f (--t)) w = fourier\_transform f (--w) \hspace{-1.0cm}   }}
 \end{array}$}}    \\ \hline

%%%%%%%%%%%%%%%%%%%%%%%%%%%%%%%%%%%%%%%%%%%

    %Line 7

\multicolumn{2}{c}{\textbf{First-order Differentiation}} \\ \hline

    \multicolumn{1}{l}{ $ \mathcal{F} [\dfrac{d}{dt}f(t) ] = i \omega F(\omega) $  }   &

    \multicolumn{1}{l}{{$\begin{array} {lcl} \textup{\texttt{\hspace{-0.3cm}$\vdash$ $\forall$ f w. fourier\_exists f $\wedge$  }} \\
\textup{\texttt{$\mathtt{\ }$\hspace{0.2cm} fourier\_exists ($\lambda$t. vector\_derivative f (at t)) $\wedge$  }}  \\
\textup{\texttt{$\mathtt{\ }$\hspace{0.2cm} ($\forall$t. f differentiable at t) $\wedge$ }}  \\
\textup{\texttt{$\mathtt{\ }$\hspace{0.2cm} (($\lambda$t. f $\mathtt{\overline{t}}$) $\rightarrow$ vec 0) at\_posinfinity $\wedge$  }}  \\
\textup{\texttt{$\mathtt{\ }$\hspace{0.2cm} (($\lambda$t. f $\mathtt{\overline{t}}$) $\rightarrow$ vec 0) at\_neginfinity  }}  \\
\textup{\texttt{$\mathtt{\ }$\hspace{-0.2cm} $\Rightarrow$ fourier\_transform ($\lambda$t. vector\_derivative f (at t)) w = \hspace{-1.0cm} }}  \\
\textup{\texttt{$\mathtt{\ }$\hspace{2.80cm} ii $\ast$ Cx w $\ast$ fourier\_transform f w  }}
 \end{array}$}}    \\ \hline

%%%%%%%%%%%%%%%%%%%%%%%%%%%%%%%%%%%%%%%%%%%

    %Line 8

\multicolumn{2}{c}{\textbf{Higher-order Differentiation}} \\ \hline

    \multicolumn{1}{l}{$ \mathcal{F} [\dfrac{d^n}{{dt}^n}f(t)] = (i \omega)^n F(\omega)$ }   &

    \multicolumn{1}{l}{{$\begin{array} {lcl} \textup{\texttt{\hspace{-0.3cm}$\vdash$ $\forall$ f w n. fourier\_exists\_higher\_deriv n f $\wedge$  }} \\
\textup{\texttt{$\mathtt{\ }$\hspace{-0.2cm} ($\forall$t. differentiable\_higher\_derivative n f t) $\wedge$   }}  \\
\textup{\texttt{$\mathtt{\ }$\hspace{-0.2cm} ($\forall$k. k < n $\Rightarrow$ }}  \\
\textup{\texttt{$\mathtt{\ }$\hspace{0.0cm} (($\lambda$t. higher\_vector\_derivative k f $\mathtt{\overline{t}}$) $\rightarrow$ vec 0)  }}  \\
\textup{\texttt{$\mathtt{\ }$\hspace{1.2cm} at\_posinfinity) $\wedge$  }}  \\
\textup{\texttt{$\mathtt{\ }$\hspace{-0.2cm} ($\forall$k. k < n $\Rightarrow$  }}  \\
\textup{\texttt{$\mathtt{\ }$\hspace{0.0cm} (($\lambda$t. higher\_vector\_derivative k f $\mathtt{\overline{t}}$) $\rightarrow$ vec 0)  }} \\
\textup{\texttt{$\mathtt{\ }$\hspace{1.2cm} at\_neginfinity)  }} \\
\textup{\texttt{$\mathtt{\ }$\hspace{-0.5cm} $\Rightarrow$ fourier\_transform ($\lambda$t. higher\_vector\_derivative n f t) w =   \hspace{-1.0cm} }} \\
\textup{\texttt{$\mathtt{\ }$\hspace{2.14cm} (ii $\ast$ Cx w) pow n $\ast$ fourier\_transform f w  }}
 \end{array}$}}    \\ \hline

    %Line 9

\multicolumn{2}{c}{\textbf{Area Under a Function}} \\ \hline

    \multicolumn{1}{l}{    $ \int_{-\infty}^{\infty} {f (t)} dt = F(0) $      }   &

    \multicolumn{1}{l}{{$\begin{array} {lcl} \textup{\texttt{\hspace{-0.3cm}$\vdash$ $\forall$ f. fourier\_exists f $\Rightarrow$  }} \\
\textup{\texttt{$\mathtt{\ }$\hspace{-0.3cm}  integral UNIV f = fourier\_transform f (\&0)  }}
 \end{array}$}}    \\ \hline

%%%%%%%%%%%%%%%%%%%%%%%%%%%%%%%%%%%%%%%%%%%

    \end{longtable}

\end{scriptsize}

%-----------------------------------------------------------------------------------------------------------------------------------------------------------------------
%-----------------------------------------------------------------------------------------------------------------------------------------------------------------------

The first property is \textit{linearity}, which is frequently used for the analysis of systems having composition of subsystems and accept different scaled inputs.

Next, we verified the \textit{time shifting} property, which is usually used to evaluate the Fourier transform of the function $f$ that is shifted over some constant value of time. The time shifting of the function $f$ can be towards the left of the origin of the time axis (time advance) or towards the right side of the origin of the time-axis (time delay). Its mathematical and the formalized form is given in Table~\ref{TAB:properties_of_Fourier_transform}.

The \textit{frequency shifting} property of Fourier transform is usually used to evaluate the Fourier transform of multiplication of the function $f$ with the exponential function. It basically shifts the frequency domain representation of $f$ to a certain portion of the frequency spectrum, which is desired for the corresponding frequency analysis. Similar to the time shifting, the frequency shifting is of two types. The frequency right shifting (frequency delay) shifts the frequency signal to the right on the frequency axis
and the frequency left shifting (frequency advance) shifts the frequency signal to the left on the frequency axis. The mathematical and the formalized forms of both versions of the frequency shifting are given in Table~\ref{TAB:properties_of_Fourier_transform}.

The next entry in Table~\ref{TAB:properties_of_Fourier_transform} presents a variant of frequency shifting, called the \textit{modulation} property, which is usually used to evaluate the Fourier transform of multiplication of the function $f$ with the cosine and sine functions. This property forms the basis of the Amplitude Modulation (AM) in communication systems. The multiplication of the sinusoidal functions (carrier signals) with the function $f$ in time-domain shifts the frequency components to the portion of the frequency spectrum that is desired for a particular signal transmission.

Next, we verified the \textit{time scaling} property of Fourier transform of a function $ f $, as given in Table~\ref{TAB:properties_of_Fourier_transform}. Here $ a: \mathds{R} $ is an arbitrary constant. If $|a| < 1$, then the function $f (at)$ represents the function $f$ compressed by a factor of $a$ and its resulting frequency spectrum will be expanded by the same factor. Similarly, in the case of $|a| > 1$, the function $f (at)$ is expanded by the factor $a$ and its corresponding frequency spectrum will be compressed by the same factor.
The next property is the \textit{time reversal} property, which is a special case of time scaling property, under the condition $a = -1$.

The Fourier transform of the \textit {differential of a function} $ f $ is a very important property that enables us to evaluate the frequency spectrum of the derivative of a function $ f $ using the Fourier transform of $ f $. Its mathematical and formalized form is presented in Table~\ref{TAB:properties_of_Fourier_transform}. In its formalized form, the first two assumptions ensure that the Fourier transforms of the function $\texttt{f}$ and its derivative $ \frac{df}{dt} $ exist. The third assumption models the condition that the function $ \texttt{f} $ is differentiable at every $\texttt{t} \ \epsilon \ \mathds{R}$ . The last two assumptions represent the condition that $ \lim\limits_{t \to \pm\infty} {f (t)} = 0 $. Finally, the conclusion provides the Fourier transform of the first order derivative of the given function.
The proof of this property involves a significant amount of arithmetic reasoning along with the integration by parts and the fact $ f(t)e^{-i \omega t} {\mid}_{-\infty}^{\infty} = ( \lim\limits_{B \to \infty} {f (B)e^{-i \omega B}}  - \lim\limits_{A \to - \infty} {f (A)e^{-i \omega A}} ) = 0 $
and the integrability of the Fourier integrand on the positive and negative real lines.

The next property is the \textit{Fourier transform of a n-times continuously differentiable function} $ f $, which is the foremost foundational property for analysing higher-order differential equations based on the Fourier transform. In its formalized form, the first assumption ensures the Fourier transform existence of $\texttt{f}$ and its first $\texttt{n}$ higher-order derivatives. Similarly, the second assumption ensures the differentiability of $\texttt{f}$ and its first $\texttt{n}$ higher-order derivatives on $ \texttt{t} \ \epsilon \ \mathds{R} $. The next two assumptions model the condition $ \lim\limits_{t \to \pm\infty} {f^{(k)} (t)} = 0 $ for each $ k = 0,1,2,...,n - 1 $, where $ f^{(k)} $ denotes the $k^{th}$ derivative of $\texttt{f}$ and $ f^{(0)} = \texttt{f} $. Finally, the conclusion is the Fourier transform of $n^{th}$ order derivative of the function. Its proof is mainly based on induction on variable $\texttt{n}$ along with Fourier transform of the first order derivative of the given function.
The Fourier transform can be used to evaluate the \textit{area under a function} $f$, as given in the final entry of Table~\ref{TAB:properties_of_Fourier_transform}.

\subsection{Relationship with Various Transforms}

This section presents the relationship of Fourier transform with various transforms, which include Fourier Cosine, Fourier Sine and Laplace transforms.

\subsubsection{Relationship with Fourier Cosine and Fourier Sine Transforms}\label{SUBSEC:relation_fourier_consine_sine}

The Fourier transform of the even and odd function enables us to relate the Fourier transform to Fourier Cosine and Fourier Sine transforms. The Fourier Cosine transform is mathematically expressed by the following indefinite integral:

\begin{equation}\label{EQ:fourier_cosine}
F_c(\omega) = \int_{-\infty}^{+\infty} {f(t)cos(\omega t)} dt
\end{equation}

\squeezedown

If the input function is an even function, i.e., $ f(-t) = f(t)$ for all $ t \ \epsilon \ \mathds{R} $, then its Fourier transform is equal to its Fourier Cosine transform.
%~\citep{beerends2003fourier}.

We verify the even function property as the following theorem:

%\squeezedown

%\begin{mdframed}
%\begin{flushleft}
\begin{thm}
\label{THM:fourier_even_function}
\emph{Fourier Transform of Even Function} \\{\small
\textup{\texttt{$\vdash$ $\forall$ f w. fourier\_exists f $\wedge$ ($\forall$t. f (--t) = f t) \\
$\mathtt{\ }$\hspace{3.0cm} $\Rightarrow$ fourier\_transform f w = fourier\_cosine\_transform f w
}}}
\end{thm}
%\end{flushleft}
%\end{mdframed}

%\squeezedown

\noindent In the above theorem, the two assumptions ensure the Fourier existence of $\texttt{f}$ and model the even function condition, respectively. The conclusion presents the relationship of Fourier transform to Fourier Cosine transform.

Next, the Fourier Sine transform is mathematically expressed as:

\begin{equation}\label{EQ:fourier_sine}
F_s(\omega) = \int_{-\infty}^{+\infty} {f(t)sin(\omega t)} dt
\end{equation}

\squeezedown

If the input function is an odd function, i.e., $ f(-t) = - f(t)$ for all $ t \ \epsilon \ \mathds{R} $, then its Fourier transform is equal to its Fourier Sine transform.
%~\citep{beerends2003fourier}.

The odd function property is verified in HOL-Light as the following theorem:

%\squeezedown

%\begin{mdframed}
%\begin{flushleft}
\begin{thm}
\label{THM:fourier_odd_function}
\emph{Fourier Transform of Odd Function} \\{\small
\textup{\texttt{$\vdash$ $\forall$ f w. fourier\_exists f $\wedge$ ($\forall$t. f (--t) = --f t) \\
$\mathtt{\ }$\hspace{2.5cm} $\Rightarrow$ fourier\_transform f w = --ii $\ast$ fourier\_sine\_transform f w
}}}
\end{thm}
%\end{flushleft}
%\end{mdframed}

%\squeezedown
In the above theorem, the first assumption presents the condition of the Fourier existence of the function $\texttt{f}$, whereas the second assumption models the odd function condition.

\subsubsection{Relationship with Laplace Transform}\label{SUBSEC:relation_fourier_laplace}

By restricting the complex-valued function $f:\mathds{R}^1 \rightarrow \mathds{R}^2$ and the variable $s:\mathds{R}^2$ for Laplace Transform, we can find a very important relationship between Fourier and Laplace transforms.
The Laplace transform of a function $f$ is given by the following equation:
%~\citep{beerends2003fourier}:

\begin{equation}\label{EQ:laplace_transform}
F(s) = \int_{0}^{\infty} {f(t)e^{-s t}} dt, \ s \ \epsilon \ \mathds{C}
\end{equation}

\squeezedown

A formalized form of the Laplace transform is as follows~\citep{taqdees2013formalization}:

%\squeezedown

%\begin{mdframed}
%\begin{flushleft}
\begin{defn}
\label{DEF:laplace_transform}
\emph{Laplace Transform} \\{\small
\textup{\texttt{$\vdash$ $\forall$ s f. laplace\_transform f s =  \\
$\mathtt{\ }$\hspace{1.2cm} lim at\_posinfinity ($\lambda$b. integral (interval [$\mathtt{\overline{\&0}}$, $\mathtt{\overline{b}}$]) \\
$\mathtt{\ }$\hspace{6.75cm} ($\lambda$t. cexp (--(s $\ast$ Cx $\mathtt{\underline{t}}$)) $\ast$ f t))
}}}
\end{defn}
%\end{flushleft}
%\end{mdframed}

%\squeezedown

The Laplace transform of a function $f$ exists, if the function $\mathtt{f}$ is piecewise smooth and of exponential order on the positive real line.
The existence of the Laplace transform has been formally defined as follows~\citep{taqdees2013formalization,rashid2017tmformalization}:

%\squeezedown

%\begin{mdframed}
%\begin{flushleft}
\begin{defn}
\label{DEF:laplace_existence}
\emph{Laplace Exists} \\{\small
\textup{\texttt{$\vdash$ $\forall$ s f. laplace\_exists f s $\Leftrightarrow$ \\
$\mathtt{\ }$\hspace{1.2cm} ($\forall$ b. f piecewise\_differentiable\_on interval [$\mathtt{\overline{\&0}}$, $\mathtt{\overline{b}}$]) $\wedge$ \\
$\mathtt{\ }$\hspace{1.2cm} ($\exists$ M a. Re s > $\mathtt{\underline{a}}$ $\wedge$ exp\_order f M a)
}}}
\end{defn}
%\end{flushleft}
%\end{mdframed}

%\squeezedown

The function $\texttt{exp\_order}$ in the above definition has been formally defined as~\citep{taqdees2013formalization,rashid2017tmformalization}:

%\squeezedown

%\begin{mdframed}
%\begin{flushleft}
\begin{defn}
\label{DEF:exp_order_condition}
\emph{Exponential Order Function} \\{\small
\textup{\texttt{$\vdash$ $\forall$ f M a. exp\_order f M a $\Leftrightarrow$ \&0 < M $\wedge$ \\
$\mathtt{\ }$\hspace{3.0cm} ($\forall$ t. \&0 <= t $\Rightarrow$ norm (f $\mathtt{\overline{t}}$) <= M $\ast$ exp ($\mathtt{\underline{a}}$ $\ast$ t))
}}}
\end{defn}
%\end{flushleft}
%\end{mdframed}

%\squeezedown

If the function $f$ is causal, i.e., $f (t) = 0$ for all $t < 0$ and the real part of Laplace variable $ \mathtt{s: R^2} $ is zero, i.e., $ \textit{Re s = 0} $, then the Fourier transform of function $f$ is equal to Laplace transform, i.e., $ {(\mathcal{F} f)(Im \ s) = (\mathcal{L} f)(s)\mid_{\textit{Re s = 0}}} $~\citep{thomas2016analysis}.

The above relationship is verified in HOL-Light as follow:

%\squeezedown

%\begin{mdframed}
%\begin{flushleft}
\begin{thm}
\label{THM:relation_fourier_laplace}
\emph{Relationship with Laplace Transform} \\{\small
\textup{\texttt{$\vdash$ $\forall$ f s. laplace\_exists f s $\wedge$ \\
$\mathtt{\ }$\hspace{0.2cm} ($\forall$t. t IN \{t | $\mathtt{\underline{t}}$ <= \&0\} $\Rightarrow$ f t = vec 0) $\wedge$ ($\forall$t. Re s = \&0) \\
$\mathtt{\ }$\hspace{3.8cm} $\Rightarrow$ fourier\_transform f (Im s) = laplace\_transform f s
}}}
\end{thm}
%\end{flushleft}
%\end{mdframed}

%\squeezedown

The first assumption of above theorem ensure the existence of the Laplace transform. The next two assumptions ensure that $\texttt{f}$ is a causal function and the real part of the Laplace variable $\texttt{s}$ is zero. The proof of the above theorem is mainly based on the integrability of the Fourier integrand on the positive and negative real lines, properties of the complex exponential,
%Lemmas \ref{THM:prop_01_INTEGRABLE_LIM_AT_POSINFINITY}, \ref{THM:prop_08_integrand_integ_pos_real_line} and \ref{THM:prop_08_integrand_integ_neg_real_line}
and the following important lemma:

%\squeezedown

%\begin{mdframed}
%\begin{flushleft}
\begin{lem}
\label{THM:laplace_alternate_representation}
\emph{Alternative Representation of Laplace Transform} \\{\small
\textup{\texttt{$\vdash$ $\forall$ f s. laplace\_exists f s $\Rightarrow$ \\
$\mathtt{\ }$\hspace{1.0cm} laplace\_transform f s =  \\
$\mathtt{\ }$\hspace{1.0cm} integral \{t | \&0 <= $\mathtt{\underline{t}}$\}  ($\lambda$t. cexp (--(s $\ast$ Cx $\mathtt{\underline{t}}$)) $\ast$ f t)
}}}
\end{lem}
%\end{flushleft}
%\end{mdframed}

%\squeezedown

\noindent The above lemma presents an alternative representation of the Laplace transform, given in Definition \ref{DEF:laplace_transform}.
This alternate representation of Laplace transform as well as the formalization of Fourier transform, given in Definition~\ref{DEF:fourier_transform}, is better than the formal definition (Definition~\ref{DEF:laplace_transform}), presented in~\citep{taqdees2013formalization}, which involves the notion of the limit. As the HOL-Light definition of the integral function implicitly encompasses infinite limits of integration, so we do not need to involve the notion of limit. Hence, this alternate representation covers the region of integration, i.e., $[0, \infty)$, as \texttt{\small{\{t | \&0 <= drop t\}}} and is equivalent to the definition of Laplace transform given by Definition~\ref{DEF:laplace_transform}.  Similarly, the region of integration for Fourier transform, i.e., $(-\infty, \infty)$ is modeled as \texttt{\small{UNIV}}.
This relationship (Theorem~\ref{THM:laplace_alternate_representation}) can facilitate the formal reasoning process of Laplace transform related properties and thus can be very useful towards the formalization of inverse Laplace transform function and verification of its associated properties.
Moreover, the formal definition of the Fourier transform presented as Definition~\ref{DEF:fourier_transform} considerably simplifies the reasoning process in the verification of its properties.

\subsection{Differential Equation}\label{SUBSEC:differential_eq_property}

Differential equations are widely used to mathematical model the complex dynamics of a continuous-time system and hence characterize the behavior of the system at each time instant.
A general linear differential equation can be mathematically expressed as follow:

\begin{equation} \label{EQ:diff_eqn_nth_order}
\begin{split}
  \textit{Differential} \ \textit{Equation} & = \sum _{k = 0}^{n} {{\alpha}_k \dfrac{d^ky}{{dt}^k}} \\
 & = {{\alpha}_n \dfrac{d^ny}{{dt}^n}} + {{\alpha}_{n-1} \dfrac{d^{n-1}y}{{dt}^{n-1}}} + ... + {{\alpha}_1 \dfrac{d^1y}{{dt}^1}} + {{\alpha}_0 y}
\end{split}
\end{equation}

\noindent where $ n $ is the order of the differential equation and $ \alpha_i $ represents the list of the constant coefficients. The Fourier transform of the above $n^{th}$-order differential equation is given by the following mathematical expression:

\begin{equation}\label{EQ:ft_diff_eqn_nth_order}
\mathcal{F} \Big( \sum _{k = 0}^{n} {{\alpha}_k \dfrac{d^ky}{{dt}^k}} \Big) = Y(\omega) \ \sum _{k = 0}^{n} {{\alpha}_k {(i \omega)}^k}
\end{equation}

We formalize the above differential equation using the following definition in HOL-Light:

%\begin{mdframed}
%\begin{flushleft}
\begin{defn}
\label{DEF:diff_eqn_nth_order_con_coef}
\emph{Differential Equation of Order $n$} \\{\small
\textup{\texttt{$\vdash$ $\forall$ n lst f t. differential\_equation n lst f t = \\
$\mathtt{\ }$\hspace{2.8cm} vsum (0..n) ($\lambda$k. EL k lst $\ast$ higher\_order\_derivative k f t)
}}}
\end{defn}
%\end{flushleft}
%\end{mdframed}

The function $\mathtt{differential\_equation}$ accepts the order of the differential equation $\texttt{n}$, a list of constant coefficients $\texttt{lst}$, a differentiable function $\texttt{f}$ and the differentiation variable $\texttt{t}$. It utilizes the functions $\mathtt{vsum \ n \ f}$ and $\mathtt{EL \ m \ lst}$, which return the vector summation $\sum_{i=0}^{n}f_i$ and the $m^{th}$ element of a list $\texttt{lst}$, respectively, to generate the differential equation corresponding to the given parameters.

Next, we verify the Fourier transform of a linear differential equation, which is expected to be the most widely used result of our formalization as depicted in Sections~\ref{SEC:formal_analysis_generic_n_order_sys} and~\ref{SEC:applications}, and is given by the following theorem in HOL-Light.

\begin{thm}
\label{THM:fourier_transform_of_diff_equation}
\emph{Fourier Transform of Differential Equation of Order $n$} \\{\small
\textup{\texttt{$\vdash$ $\forall$ f lst w n. fourier\_exists\_higher\_deriv n f $\wedge$   \\
$\mathtt{\ }$\hspace{0.2cm} ($\forall$t. differentiable\_higher\_derivative n f t) $\wedge$  \\
$\mathtt{\ }$\hspace{0.2cm} ($\forall$k. k < n $\Rightarrow$  \\
$\mathtt{\ }$\hspace{1.2cm}(($\lambda$t. higher\_vector\_derivative k f $\mathtt{\overline{t}}$) $\rightarrow$ vec 0) at\_posinfinity) $\wedge$ \\
$\mathtt{\ }$\hspace{0.2cm} ($\forall$k. k < n $\Rightarrow$  \\
$\mathtt{\ }$\hspace{1.2cm}(($\lambda$t. higher\_vector\_derivative k f $\mathtt{\overline{t}}$) $\rightarrow$ vec 0) at\_neginfinity) \\
$\mathtt{\ }$\hspace{1.8cm} $\Rightarrow$ fourier\_transform ($\lambda$t. differential\_equation n lst f t) w = \\
$\mathtt{\ }$\hspace{1.3cm} fourier\_transform f w $\ast$ vsum (0..n) ($\lambda$k. EL k lst $\ast$ (ii $\ast$ Cx w) pow k)
}}}
\end{thm}

The set of the assumptions of the above theorem is the same as that of property named \textit{Higher Order Differentiation in Time Domain}, given in Table~\ref{TAB:properties_of_Fourier_transform}. The conclusion of Theorem \ref{THM:fourier_transform_of_diff_equation} is the Fourier transform of a $n^{th}$-order linear differential equation.
The proof of above theorem is based on induction on variable $\texttt{n}$. The proof of the base case is based on simple arithmetic reasoning and the step case is discharged using Theorems \ref{THM:linearity_prop_four_exist}, \textit{linearity} and \textit{Higher Order Differentiation in Time Domain} properties, along with the following important lemma about the Fourier existence of the differential equation.

\begin{lem}
\label{THM:fourier_existence_of_diff_equation}
\emph{Fourier Existence of Differential Equation} \\{\small
\textup{\texttt{$\vdash$ $\forall$ n lst f. fourier\_exists\_higher\_deriv n f  \\
$\mathtt{\ }$\hspace{3.0cm} $\Rightarrow$ fourier\_exists ($\lambda$t. differential\_equation n lst f t)
}}}
\end{lem}

\section{Fourier Transform of Some Commonly Used Functions} \label{SEC:fourier_trans_comm_used_funs}

In this section, we present Fourier transform of some functions that are commonly used for the analysis of the physical and engineering systems in various domains, i.e., signal processing, analog circuits, optical systems and communication systems etc.
\subsection{Fourier Transform of a Rectangular Pulse}

The rectangular pulse is characterized by having a constant value over a range of values on the time axis. It is represented by the following mathematical expression.

  \begin{equation}\label{EQ:rect_pulse}
    f (t) =
    \begin{cases}
      1, & |t| \leq T_1 \\
      0, & \text{otherwise}
    \end{cases}
  \end{equation}

It has a constant value $1$ inside the interval $[-T_1, T_1]$, whereas it is $0$ outside this interval over the whole real line. %This pulse signal is extensively used in signal processing and communication systems.
For the value of $T_1 = 0.5$, it is known as the unit gate function. The Fourier transform of the rectangular pulse in given by the following equation.

\begin{equation} \label{EQ:fourier_rect_pulse}
\begin{split}
  F (\omega) & = 2 \textit{T}_1 \textit{sinc} (\omega \textit{T}_1)  \\
 & = 2 \textit{T}_1 \dfrac{\textit{sin} (\omega \textit{T}_1)}{\omega \textit{T}_1}
\end{split}
\end{equation}

%\begin{equation}\label{EQ:fourier_rect_pulse}
%F (\omega) = 2 \textit{T}1 \textit{sinc} (\omega \textit{T}1) = 2 \textit{T}1 \dfrac{\textit{sin} (\omega \textit{T}1)}{\omega \textit{T}1}
%\end{equation}

\noindent where $\textit{sinc} (\omega \textit{T}_1)$ is the sinc function, which is the multiplication of a sinusoidal function $sin (\omega \textit{T}_1)$ with the monotonically decreasing function $\dfrac{1}{\omega \textit{T}_1}$, which makes it a continuously decreasing sinusoidal function. It approaches to value $0$ at both endpoints of the axis. i.e., $-\infty$ and $+\infty$. It is also known as the interpolation or filtering function. We model the rectangular pulse and the sinc function using the following HOL-Light functions:

\begin{defn}
\label{DEF:rect_pulse}
\emph{Rectangular Pulse} \\{\small
\textup{\texttt{$\vdash$ rect\_pulse T1 = \\
$\mathtt{\ }$\hspace{1.7cm} ($\lambda$t. if t IN \{t | --$\mathtt{\underline{T1}}$ <= $\mathtt{\underline{t}}$ $\wedge$ $\mathtt{\underline{t}}$ <= $\mathtt{\underline{T1}}$\}  \\
$\mathtt{\ }$\hspace{2.6cm} then Cx (\&1) \\
$\mathtt{\ }$\hspace{2.6cm} else Cx (\&0))
}}}
\end{defn}

\begin{defn}
\label{DEF:sinc_fun}
\emph{Sinc Function} \\{\small
\textup{\texttt{$\vdash$ sinc T1 w = csin (Cx w $\ast$ Cx $\mathtt{\underline{T1}}$) / (Cx w $\ast$ Cx $\mathtt{\underline{T1}}$)
}}}
\end{defn}

The Fourier transform of the rectangular pulse is represented as the following HOL-Light theorem:

\begin{thm}
\label{THM:fourier_rect_pulse}
\emph{Fourier Transform of Rectangular Pulse} \\{\small
\textup{\texttt{$\vdash$ $\forall$ T1 w.  \&0 < $\mathtt{\underline{T1}}$ $\wedge$ $\sim$(w = \&0)    \\
$\mathtt{\ }$\hspace{1.7cm} $\Longrightarrow$ fourier\_transform (rect\_pulse T1) w = \\
$\mathtt{\ }$\hspace{6.7cm} Cx (\&2) $\ast$ Cx $\mathtt{\underline{T1}}$ $\ast$ sinc T1 w
}}}
\end{thm}

We verified the above theorem using Definitions \ref{DEF:fourier_transform}, \ref{DEF:rect_pulse} and \ref{DEF:sinc_fun}, and the properties of the integration along with some arithmetic reasoning.
%It is further used in the formal analysis of the audio equalizer, presented in Section \ref{SEC:applications}.

\subsection{Fourier Transform of Unilateral Negative Complex Exponential}
The unilateral negative complex exponential is given by the following mathematical expression:

  \begin{equation}\label{EQ:one_sided_neg_cexp}
    f (t) =
    \begin{cases}
      0, & t < 0 \\
      e^{-ct}, & t \geq 0
    \end{cases}
  \end{equation}

\noindent where $c$ is a positive real constant, which makes the function $e^{-ct}$ an exponentially decaying function.
%It is widely used as the input signal to many physical and engineering systems.
The Fourier transform of the unilateral complex exponential function is as below:

\begin{equation}\label{EQ:fourier_one_sided_neg_exp}
F (\omega) = \dfrac{1}{c + i \omega}
\end{equation}

We formally model the unilateral negative complex exponential by the following HOL-Light function:

\begin{defn}
\label{DEF:one_sided_neg_cexp}
\emph{Unilateral Negative Complex Exponential} \\{\small
\textup{\texttt{$\vdash$ unilat\_neg\_cexp c = \\
$\mathtt{\ }$\hspace{1.7cm} ($\lambda$t. if t IN \{t | \&0 <= $\mathtt{\underline{t}}$\}  \\
$\mathtt{\ }$\hspace{2.6cm} then cexp (--Cx c $\ast$ Cx $\mathtt{\underline{t}}$) \\
$\mathtt{\ }$\hspace{2.6cm} else Cx (\&0))
}}}
\end{defn}

We verified the Fourier transform of the unilateral negative complex exponential as the following theorem.

\begin{thm}
\label{THM:fourier_one_sided_neg_exp}
\emph{Fourier Transform of Unilateral Negative Exponential} \\{\small
\textup{\texttt{$\vdash$ $\forall$ c w. \&0 < c $\wedge$ $\sim$(Cx c + ii $\ast$ Cx w = Cx (\&0))    \\
$\mathtt{\ }$\hspace{1.7cm} $\Longrightarrow$ fourier\_transform (unilat\_neg\_cexp c) w = \\
$\mathtt{\ }$\hspace{6.7cm} Cx (\&1) / (Cx c + ii $\ast$ Cx w)
}}}
\end{thm}

%The verification process of the above theorem is based on Definitions \ref{DEF:fourier_transform} and \ref{DEF:one_sided_neg_cexp}, and properties of integration along with some arithmetic reasoning.

\subsection{Fourier Transform of Bilateral Complex Exponential}

The Fourier transform of the bilateral complex exponential is given by the following mathematical equation:

\begin{equation}\label{EQ:fourier_bilateral_exp}
\mathcal{F} [e^{-|t|}]  = \dfrac{2}{1 + \omega^2}
\end{equation}

We verified its Fourier transform as the following theorem:

\begin{thm}
\label{THM:fourier_bilateral_exp}
\emph{Fourier Transform of Bilateral Complex Exponential} \\{\small
\textup{\texttt{$\vdash$ $\forall$ w. $\sim$(Cx (\&1) - ii $\ast$ Cx w = Cx (\&0)) $\wedge$   \\
$\mathtt{\ }$\hspace{0.88cm} $\sim$(Cx (\&1) + ii $\ast$ Cx w = Cx (\&0))   \\
$\mathtt{\ }$\hspace{1.7cm} $\Longrightarrow$ fourier\_transform ($\lambda$t. cexp (--Cx (abs $\mathtt{\underline{t}}$))) w \\
$\mathtt{\ }$\hspace{6.7cm} Cx (\&2) / (Cx (\&1) + Cx w pow 2)
}}}
\end{thm}

The verification process of the above theorem starts by rewriting with the definition of Fourier transform. Next, we split the region of integration, i.e., $(-\infty, +\infty)$, into $(-\infty, 0]$ and $[0, +\infty)$, as a result, we obtain two integrals with the same integrand and the respective regions of integration. We rewrite the resultant subgoal with the definition of absolute value of a real number to replace its value. i.e., $|t| = -t$ in interval $(-\infty, 0]$ and $|t| = t$ in interval $[0, +\infty)$. Next, these two integrals are evaluated using
%Theorem \ref{THM:prop_01_INTEGRABLE_LIM_AT_POSINFINITY} and
the properties of integration along with the complex arithmetic reasoning to conclude the proof of Theorem~\ref{THM:fourier_bilateral_exp}.

\subsection{Fourier Transform of Finite Duration Sinusoidal Tone-burst}

The sinusoidal tone-burst occurring for a finite duration $-T_1$ to $T_1$ is mathematically defined as follows:

  \begin{equation}\label{EQ:sine_tone_burst}
    f (t) =
    \begin{cases}
      sin \omega_0 t, & |t| \leq T_1 \\
      0, & \text{otherwise}
    \end{cases}
  \end{equation}

The Fourier transform of the above sinusoidal tone-burst is given by the following mathematical expression:

\begin{equation}\label{EQ:fourier_sine_tone_burst}
F (\omega) = - i T_1 \{ \textit{sinc} ((\omega - \omega_0) \textit{T}_1) - \textit{sinc} ((\omega + \omega_0) \textit{T}_1) \}
\end{equation}

The frequency spectrum corresponding to the above sine wave is inversely proportional to its duration $2T_1$. i.e., with the increase in the duration $2T_1$, it results into a narrower frequency line spectrum and vice versa.

We defined Equation \ref{EQ:sine_tone_burst} as the following HOL-Light function:

\begin{defn}
\label{DEF:sine_tone_burst}
\emph{Finite Duration Sinusoidal Tone-burst} \\{\small
\textup{\texttt{$\vdash$  $\forall$ T1 w0.  sine\_tone\_burst T1 w0 = \\
$\mathtt{\ }$\hspace{3.7cm} ($\lambda$t. csin (Cx w0 $\ast$ Cx $\mathtt{\underline{t}}$) $\ast$ rect\_pulse T1 t)
}}}
\end{defn}

The Fourier transform of the above sinusoidal tone-burst is represented as the following theorem:

\begin{thm}
\label{THM:fourier_sine_tone_burst}
\emph{Fourier Transform of Finite Duration Sinusoidal Tone-burst} \\{\small
\textup{\texttt{$\vdash$ $\forall$ T1 w w0. \&0 < $\mathtt{\underline{T1}}$ $\wedge$     \\
$\mathtt{\ }$\hspace{1.88cm} $\sim$(w - w0 = \&0) $\wedge$ $\sim$(w + w0 = \&0) \\
$\mathtt{\ }$\hspace{2.4cm} $\Longrightarrow$ fourier\_transform (sine\_tone\_burst T1 w0) w = \\
$\mathtt{\ }$\hspace{2.8cm} --ii $\ast$ Cx $\mathtt{\underline{T1}}$ $\ast$ (sinc T1 (w - w0) - sinc T1 (w + w0))
}}}
\end{thm}

We start the verification process of the above theorem by rewriting with Definitions \ref{DEF:fourier_transform}, \ref{DEF:sine_tone_burst} and \ref{DEF:rect_pulse}, and the definition of complex sine $\texttt{csin}$ (Definition \ref{DEF:exp_ccos_csine}), which results into a subgoal having vector integral of the linear combination of the $\texttt{cexp}$ functions over interval $[-T_1, T_1]$, which is verified based on the properties of integration along with the arithmetic reasoning.

\subsection{Fourier Transform of Damped Unilateral Sinusoidal Function}

The damped unilateral sinusoidal function is basically a product of a  decaying complex exponential function with the periodic sinusoidal function and is given as below:

  \begin{equation}\label{EQ:damped_one_sided_sine}
    f (t) =
    \begin{cases}
      0, & t < 0 \\
      e^{-ct} sin \omega_0 t , & t \geq 0
    \end{cases}
  \end{equation}

\noindent where $c$ is a positive real constant. The Fourier transform of the damped unilateral sinusoidal function is represented by the following equation:

\begin{equation}\label{EQ:fourier_damped_one_sided_sine}
F (\omega) = \dfrac{w_0}{(c + i \omega)^2 + {w_0}^2}
\end{equation}

We formalized Equation \ref{EQ:damped_one_sided_sine} as the following HOL-Light function:

\begin{defn}
\label{DEF:damped_one_sided_sine}
\emph{Damped Unilateral Sinusoidal Function} \\{\small
\textup{\texttt{$\vdash$  damped\_unilat\_sine c w0 = \\
$\mathtt{\ }$\hspace{1.5cm} ($\lambda$t. if t IN \{t | \&0 <= $\mathtt{\underline{t}}$\} \\
$\mathtt{\ }$\hspace{2.5cm} then cexp (--Cx c $\ast$ Cx $\mathtt{\underline{t}}$) $\ast$ csin (Cx w0 $\ast$ Cx $\mathtt{\underline{t}}$)  \\
$\mathtt{\ }$\hspace{2.5cm} else Cx (\&0))
}}}
\end{defn}

Its Fourier transform has been verified as the following theorem:

\begin{thm}
\label{THM:damped_one_sided_sine}
\emph{Fourier Transform of Damped Unilateral Sinusoidal Function} \\{\small
\textup{\texttt{$\vdash$ $\forall$ c w w0. \&0 < c $\wedge$     \\
$\mathtt{\ }$\hspace{1.70cm} $\sim$(Cx c + ii $\ast$ Cx (w - w0) = Cx (\&0)) $\wedge$ \\
$\mathtt{\ }$\hspace{1.70cm} $\sim$(Cx c + ii $\ast$ Cx (w + w0) = Cx (\&0)) $\wedge$ \\
$\mathtt{\ }$\hspace{1.70cm} $\sim$((Cx c + ii $\ast$ Cx w) pow 2 + Cx w0 pow 2 = Cx (\&0)) \\
$\mathtt{\ }$$\mathtt{\ }$\hspace{2.8cm} $\Longrightarrow$ fourier\_transform (damped\_unilat\_sine c w0) w = \\
$\mathtt{\ }$\hspace{5.0cm} Cx w0 / ((Cx c + ii $\ast$ Cx w) pow 2 + Cx w0 pow 2)
}}}
\end{thm}

The proof process of the above theorem involves Definitions \ref{DEF:fourier_transform} and \ref{DEF:damped_one_sided_sine} along with some properties about the
%Here, after applying some complex arithmetic manipulations, we obtain a subgoal containing the two
complex exponential functions.
%. Next, we applied the reasoning process of Theorem \ref{THM:fourier_one_sided_neg_exp} on each of the $\mathtt{cexp}$ functions, to conclude with the proof of Theorem \ref{THM:damped_one_sided_sine}.

This completes our formalization of Fourier transform in HOL-Light.
The source code of our formalization is available for download~\citep{adnan16contsystemfouriertrans} and can be utilized for further development and the analysis of continuous-time systems.

\section{Formal Analysis of a Generic n-order System} \label{SEC:formal_analysis_generic_n_order_sys}
\begin{figure}[ht!]
\centering
\scalebox{0.30}
{\includegraphics[trim={0 0.0cm 0 0.0cm},clip]{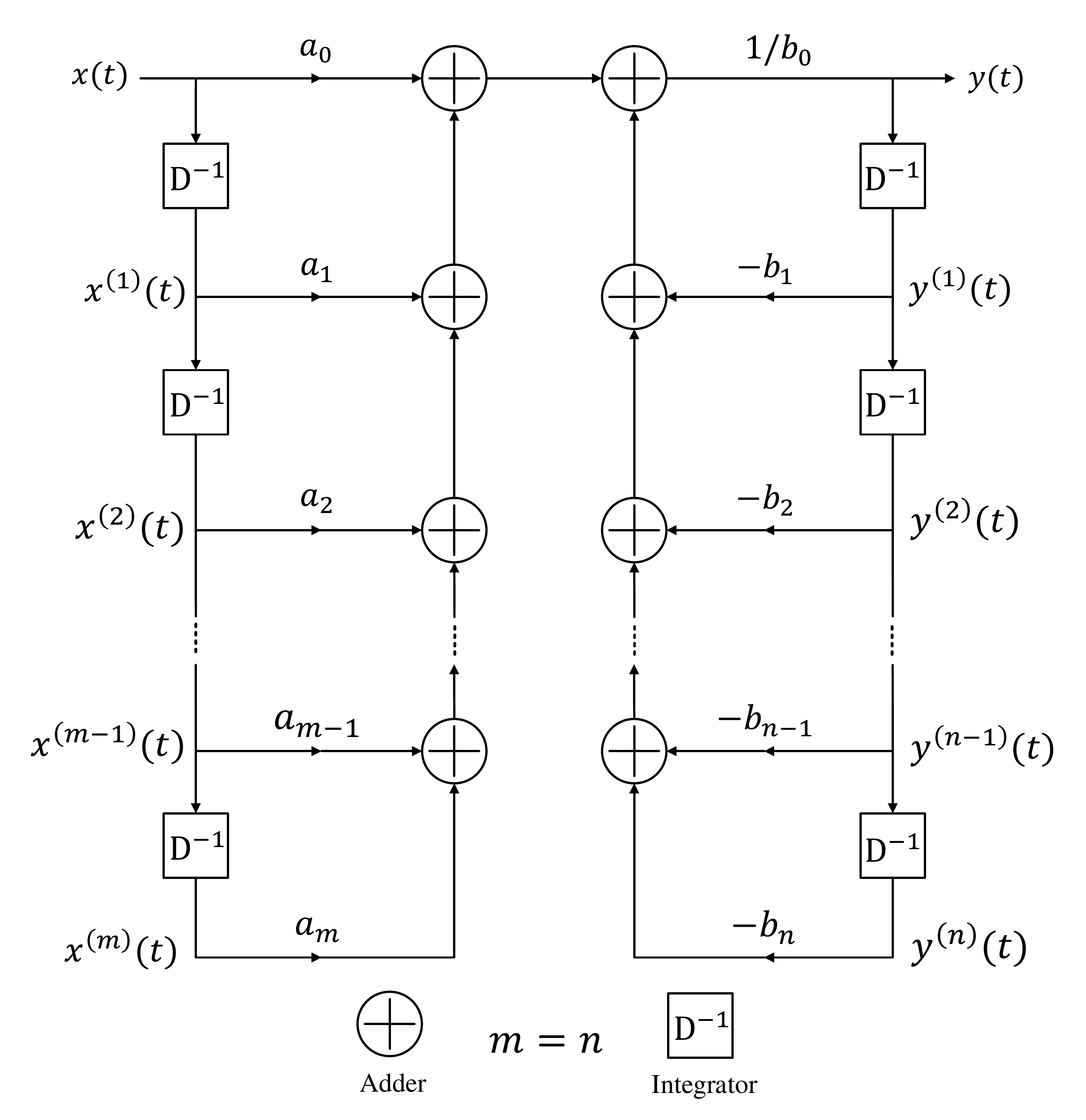}}
\caption{Block Diagram Representation of a Generic $n$-order System}
\label{FIG:nth_order_system}
\end{figure}

In this section, we present the formal modeling and the frequency response analysis of a generic n-order Linear Time Invariant (LTI) system.
A generic n-order LTI system~\citep{adams2012continuous} presents a relationship between an input signal $x(t)$ and the output signal $y(t)$ and its dynamics are modeled using a higher-order differential equation. Due to the generic nature of its differential equation based model and its corresponding frequency response analysis, we can use it for the modeling and frequency analysis of any real-world application, which certainly eases out the formal reasoning based analysis of these system and is illustrated in the next section. Fig.~\ref{FIG:nth_order_system} provides the block diagram representation of a generic $n$-order system, which is primarily composed of
the addition, the scalar multiplication and the integration operations~\citep{girod2001signals}.

The generalized linear differential equation, with constant coefficient, describing the input-output relationship for this generic $n$-order system is mathematically expressed as~\citep{adams2012continuous}:

\begin{equation}\label{EQ:diff_eqn_nth_order_LTI_sys}
 \sum _{k = 0}^{n} {{\beta}_k \dfrac{d^k}{{dt}^k} y(t)} = \sum _{k = 0}^{m} {{\alpha}_k \dfrac{d^k}{{dt}^k} x(t)}, \ \ \ \  m \leq n
\end{equation}

\noindent where $y(t)$ in the above equation is the output and $x(t)$ is the input to the system. The constants $\alpha_k$ and $\beta_k$ are the coefficients of the input and the output differentials of order $k$, respectively. The greatest index $n$ of the non-zero coefficient $\beta_n$ determines the order of the underlying system. The corresponding frequency response of the system is given by the following mathematical expression:

\begin{equation}\label{EQ:freq_res_nth_order_LTI_sys}
\dfrac{Y(\omega)}{X(\omega)} = \dfrac{\sum_{k = 0}^{m} {\alpha_k (i\omega)^k}}{\sum_{k = 0}^{n} {\beta_k (i\omega)^k}}
\end{equation}

In order to verify the above frequency response of the given system, we first model the corresponding differential equation %(Definition \ref{DEF:diff_eqn_nth_order_con_coef})
as the following HOL-Light function:

\begin{defn}
\label{DEF:diff_eqn_nth_order_lti_system}
\emph{Differential Equation of $n$-order LTI System} \\{\small
\textup{\texttt{$\vdash$ $\forall$ n outlst y m inlst x t. diff\_eq\_n\_order\_sys m n inlst outlst x y t     $\Leftrightarrow$ \\
$\mathtt{\ }$\hspace{1.2cm} differential\_equation n outlst y t = differential\_equation m inlst x t
}}}
\end{defn}

Next, we verified the frequency response, given in Equation~\ref{EQ:freq_res_nth_order_LTI_sys}, of the generic n-order system as the following HOL-Light theorem.

\begin{thm}
\label{THM:freq_response_nth_order_lti_system}
\emph{Frequency Response of $n$-order LTI System} \\{\small
\textup{\texttt{$\vdash$ $\forall$ y x m n inlst outlst w.   \\
$\mathtt{\ }$\hspace{0.0cm}  ($\forall$t. differentiable\_higher\_derivative n y t) $\wedge$  \\
$\mathtt{\ }$\hspace{0.0cm}  ($\forall$t. differentiable\_higher\_derivative m x t) $\wedge$  \\
$\mathtt{\ }$\hspace{0.0cm}  fourier\_exists\_higher\_deriv n y $\wedge$ fourier\_exists\_higher\_deriv m x $\wedge$  \\
$\mathtt{\ }$\hspace{0.0cm}   ($\forall$k. k < n $\Rightarrow$  \\
$\mathtt{\ }$\hspace{1.0cm}(($\lambda$t. higher\_vector\_derivative k y $\mathtt{\overline{t}}$) $\rightarrow$ vec 0) at\_posinfinity) $\wedge$ \\
$\mathtt{\ }$\hspace{0.0cm} ($\forall$k. k < n $\Rightarrow$  \\
$\mathtt{\ }$\hspace{1.0cm}(($\lambda$t. higher\_vector\_derivative k y $\mathtt{\overline{t}}$) $\rightarrow$ vec 0) at\_neginfinity) $\wedge$ \\
$\mathtt{\ }$\hspace{0.0cm}  ($\forall$k. k < m $\Rightarrow$  \\
$\mathtt{\ }$\hspace{1.0cm}(($\lambda$t. higher\_vector\_derivative k x $\mathtt{\overline{t}}$) $\rightarrow$ vec 0) at\_posinfinity) $\wedge$ \\
$\mathtt{\ }$\hspace{0.0cm}    ($\forall$k. k < m $\Rightarrow$  \\
$\mathtt{\ }$\hspace{1.0cm}(($\lambda$t. higher\_vector\_derivative k x $\mathtt{\overline{t}}$) $\rightarrow$ vec 0) at\_neginfinity) $\wedge$ \\
$\mathtt{\ }$\hspace{0.0cm}  ($\forall$t. diff\_eq\_n\_order\_sys m n inlst outlst x y t) $\wedge$  \\
$\mathtt{\ }$\hspace{0.0cm}   $\sim$(fourier\_transform x w = Cx (\&0)) $\wedge$ \\
$\mathtt{\ }$\hspace{0.0cm}   $\sim$(vsum (0..n) ($\lambda$t. Cx (EL t outlst) $\ast$ (ii $\ast$ Cx w) pow t) = Cx (\&0))  \\
$\mathtt{\ }$\hspace{2.2cm} $\Rightarrow$ fourier\_transform y w / fourier\_transform x w =  \\
$\mathtt{\ }$\hspace{3.05cm} vsum (0..m) ($\lambda$t. Cx (EL t inlst) $\ast$  (ii $\ast$  Cx w) pow t) / \\
$\mathtt{\ }$\hspace{3.05cm} vsum (0..n) ($\lambda$t. Cx (EL t outlst) $\ast$  (ii $\ast$  Cx w) pow t)
}}}
\end{thm}

The first two assumptions ensure that the functions $\texttt{y}$ and $\texttt{x}$ are differentiable up to the $n^{th}$ and $m^{th}$ order, respectively.
The next assumption represents the Fourier transform existence condition upto the $n^{th}$ order derivatives of function $\texttt{y}$.
Similarly, the next assumption ensures that the Fourier transform exists up to the $m^{th}$ order derivative of the function $\texttt{x}$. The next two assumptions represent the condition $ \lim\limits_{t \to \pm\infty} {y^{(k)} (t)} = 0 $ for all $ k = 0, 1, ... , n - 1 $, i.e., $ \lim\limits_{t \to \pm\infty}  {y^{(n - 1)} (t)} = 0 $, ... , $ \lim\limits_{t \to \pm\infty}  {y^{(0)} (t)} = \lim\limits_{t \to \pm\infty}  y(t) = 0 $, where $y^{(k)}(t)$ is the $k^{th}$ derivative of $\texttt{y}$ with respect to $\texttt{t}$. The next two assumptions provide the condition $ \lim\limits_{t \to \pm\infty} {x^{(k)} (t)} = 0 $ for each $ k = 0, 1, ... , m - 1 $.
The next assumption represents the formalization of Equation \ref{EQ:diff_eqn_nth_order_LTI_sys} and the last two assumptions provide some interesting design related relationships, which must hold for constructing a reliable continuous-time system. Finally, the conclusion of the above theorem represents the frequency response given by Equation \ref{EQ:freq_res_nth_order_LTI_sys}. The proof of Theorem \ref{THM:freq_response_nth_order_lti_system} was very straightforward and mainly based on Theorem~\ref{THM:fourier_transform_of_diff_equation}, along with some arithmetic reasoning, thanks to our foundational formalization presented in the previous sections. The verification of this theorem is very useful as it greatly simplifies the verification of the frequency response of any real-world application as illustrated in the next section.

\section{Applications} \label{SEC:applications}

In this section, to illustrate the utilization of our foundational formalization for analyzing real-world continuous systems, we present a formal analysis of an audio equalizer and a MEMs accelerometer. To the best of our knowledge, these systems could not have been verified while capturing their continuous behavior in the true form by any other existing computer-based analysis technique.

\subsection{Formal Analysis of an Audio Equalizer}\label{SUBSEC:audio_equalizer}

\begin{figure}[ht!]
\centering
\scalebox{0.26}
{\includegraphics[trim={0 0.0cm 0 0.0cm},clip]{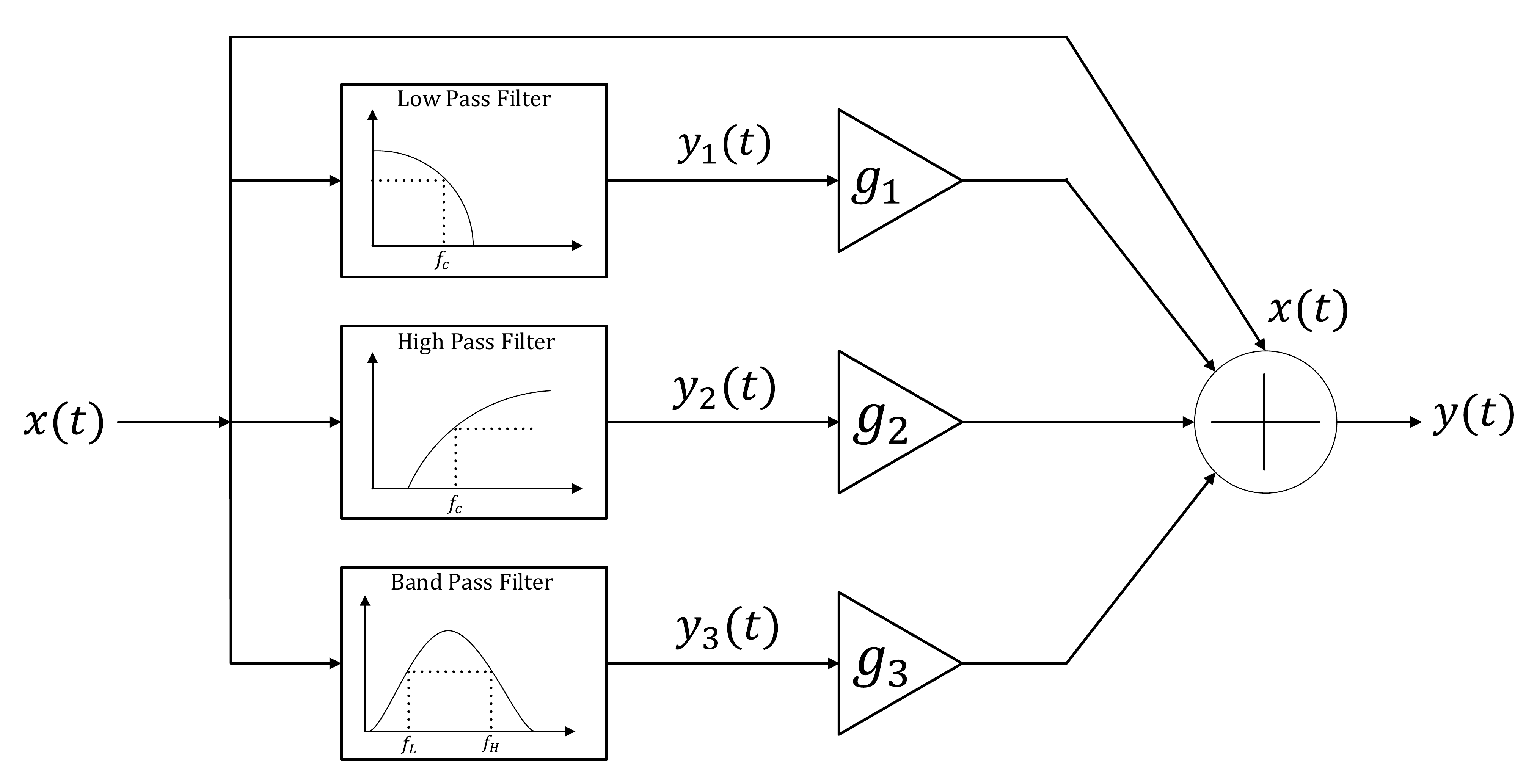}}
\caption{Block Diagram of Audio Equalizer}
\label{FIG:audio_equalizer}
\end{figure}

An audio equalizer~\citep{tan2007fundamentals} is an electronic circuit that adjusts the balance between different frequency components within an audio signal. The block diagram of a 3-channel audio equalizer is illustrated in Fig.~\ref{FIG:audio_equalizer}.

\noindent It mainly consists of three different filters, which are low-pass, high-pass and bandpass, allowing a certain range of the frequency to pass on. The low-pass filter allows the passage of signals having frequency lower than then the cutoff frequency ($\omega_c = 1/2\pi f_c$), whereas, the high-pass filter passes the signal with a frequency higher than the cutoff frequency. Whereas the bandpass filter passes the signal having frequency components in a certain range only, as shown in Figure~\ref{FIG:audio_equalizer}. After each filtering stage, some signal amplification with gain ($g_i$) is applied in order to enhance the quality of the signal. Being a major component of an audio equalizer, we verify the frequency response of each of the individual filter. Here, we only present the formal verification of the frequency response of the bandpass filter only due to space restrictions and the verification of rest of the filters can be found in the proof script~\citep{adnan16contsystemfouriertrans}.

In order to verify the frequency response of the bandpass filter, we first model its corresponding differential equation, which is given by the following HOL-Light function:

\begin{defn}
\label{DEF:diff_equation_bandpass_filter}
\emph{Differential Equation of Bandpass Filter} \\{\small
\textup{\texttt{$\vdash$  $\forall$ wc. outlst\_de\_bpf wc = [wc pow 2; \&2 $\ast$ wc; \&1]
}}} \\
{\small
\textup{\texttt{$\vdash$ $\forall$ wc. inlst\_de\_bpf wc = [\&0; wc]
}}} \\
{\small
\textup{\texttt{$\vdash$ diff\_eq\_BP\_FILTER inlst\_de\_bpf outlst\_de\_bpf x y t wc $\Leftrightarrow$ \\
$\mathtt{\ }$\hspace{2.25cm}  differential\_equation 2 (outlst\_de\_bpf wc) y t = \\
$\mathtt{\ }$\hspace{2.25cm} differential\_equation 1 (inlst\_de\_bpf wc) x t
}}}
\end{defn}

\noindent where the function \texttt{diff\_eq\_BP\_FILTER} accepts the function variables \texttt{x} and \texttt{y} and the lists of coefficients (\texttt{inlst\_de\_bpf} and \texttt{outlst\_de\_bpf}) and returns the corresponding differential equation of the bandpass filter.

Next, the frequency response of the bandpass filter is mathematically expressed as:

\begin{equation}\label{EQ:freq_res_audio_equalizer}
\begin{split}
   \frac{Y(\omega)}{X (\omega)} & = \dfrac{\omega_c}{i\omega + \omega_c} \times \dfrac{i\omega}{i\omega + \omega_c} \\
 & = \dfrac{\omega_c(i\omega)}{{(i\omega})^2 + 2\omega_c(i\omega) + (\omega_c)^2}
\end{split}
\end{equation}

\noindent% where $\omega_c$ is the cutoff frequency.
We verified the above frequency response as the following HOL-Light theorem:

\begin{thm}
\label{THM:freq_response_audio_equalizer_3}
\emph{Frequency Response of Bandpass Filter} \\{\small
\textup{\texttt{$\vdash$ $\forall$ y x w wc.  \&0 < wc  $\wedge$  \\
$\mathtt{\ }$\hspace{0.0cm} ($\forall$t. differentiable\_higher\_derivative 2 y t) $\wedge$  \\
$\mathtt{\ }$\hspace{0.0cm}  ($\forall$t. differentiable\_higher\_derivative 1 x t) $\wedge$  \\
$\mathtt{\ }$\hspace{0.0cm}  fourier\_exists\_higher\_deriv 2 y $\wedge$ fourier\_exists\_higher\_deriv 1 x $\wedge$  \\
$\mathtt{\ }$\hspace{0.0cm} ($\forall$k. k < 2 $\Rightarrow$  \\
$\mathtt{\ }$\hspace{1.0cm}(($\lambda$t. higher\_vector\_derivative k y $\mathtt{\overline{t}}$) $\rightarrow$ vec 0) at\_posinfinity) $\wedge$ \\
$\mathtt{\ }$\hspace{0.0cm} ($\forall$k. k < 2 $\Rightarrow$  \\
$\mathtt{\ }$\hspace{1.0cm}(($\lambda$t. higher\_vector\_derivative k y $\mathtt{\overline{t}}$) $\rightarrow$ vec 0) at\_neginfinity) $\wedge$ \\
$\mathtt{\ }$\hspace{0.0cm} (($\lambda$t. x $\mathtt{\overline{t}}$) $\rightarrow$ vec 0) at\_posinfinity $\wedge$ \\
$\mathtt{\ }$\hspace{0.0cm}  (($\lambda$t. x $\mathtt{\overline{t}}$) $\rightarrow$ vec 0) at\_neginfinity $\wedge$ \\
$\mathtt{\ }$\hspace{0.0cm}  ($\forall$t. diff\_eq\_BP\_FILTER inlst\_de\_bpf outlst\_de\_bpf x y t wc) $\wedge$ \\
$\mathtt{\ }$\hspace{0.0cm}   $\sim$(fourier\_transform x w = Cx (\&0)) $\wedge$ \\
$\mathtt{\ }$\hspace{0.0cm}  $\sim$((ii $\ast$ Cx w) pow 2 + Cx (\&2) $\ast$ Cx wc $\ast$ ii $\ast$ Cx w + Cx wc pow 2 = Cx (\&0)) \\
$\mathtt{\ }$\hspace{0.6cm}  $\Rightarrow$ fourier\_transform y w / fourier\_transform x w =  \\
$\mathtt{\ }$\hspace{1.5cm}  Cx wc $\ast$ ii $\ast$ Cx w /  \\
$\mathtt{\ }$\hspace{1.9cm} ((ii $\ast$ Cx w) pow 2 + Cx (\&2) $\ast$ Cx wc $\ast$ ii $\ast$ Cx w + Cx wc pow 2)
}}}
\end{thm}

The first assumption ensures that the variable corresponding to the cutoff frequency ($\texttt{wc}$) cannot be negative or zero. The next two assumptions ensure that the functions $\texttt{y}$ and $\texttt{x}$ are differentiable up to the second and first order, respectively. The next two assumptions represent the Fourier transform existence condition up to the second and first order derivatives of the functions $\texttt{y}$ and $\texttt{x}$, respectively. The next two assumptions represent the condition $ \lim\limits_{t \to \pm\infty} {y^{(k)} (t)} = 0 $ for each $k = 0, 1$.
The next two assumptions provide the condition $\lim\limits_{t \to \pm\infty}  x(t) = 0 $.
The next assumption represents the formalization of the corresponding differential equation and the last two assumptions provide some interesting design related relationships, which must hold for constructing a reliable bandpass filter. Finally, the conclusion of the above theorem represents the frequency response given by Equation \ref{EQ:freq_res_audio_equalizer}. The proof is based on Theorem~\ref{THM:freq_response_nth_order_lti_system}, along with some arithmetic reasoning.

\subsection{Formal Analysis of MEMs Accelerometer}\label{SUBSEC:mems_accele}

\begin{figure}[H]
\captionsetup[subfigure]{labelformat=empty}
\begin{subfigure}{.5\textwidth}
  \centering
  \includegraphics[width=.6\linewidth]{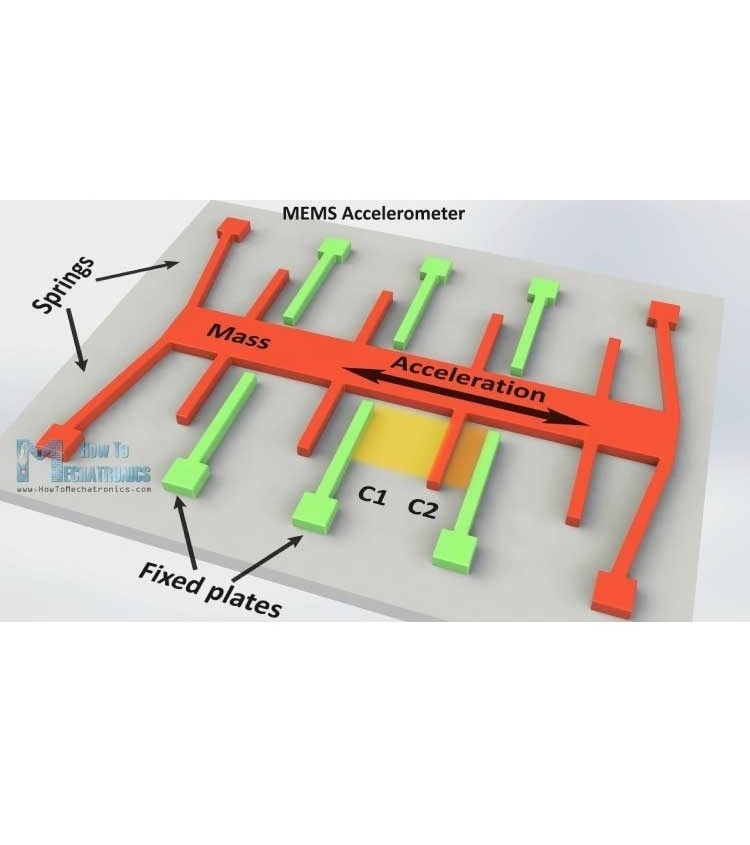}
%  \vspace*{2.0cm}
  \caption{\hspace*{3.1cm}(a)}
  \label{FIG:mems_accelerometer_diagram}
\end{subfigure}%
\begin{subfigure}{.5\textwidth}
  \centering
  \includegraphics[width=.6\linewidth]{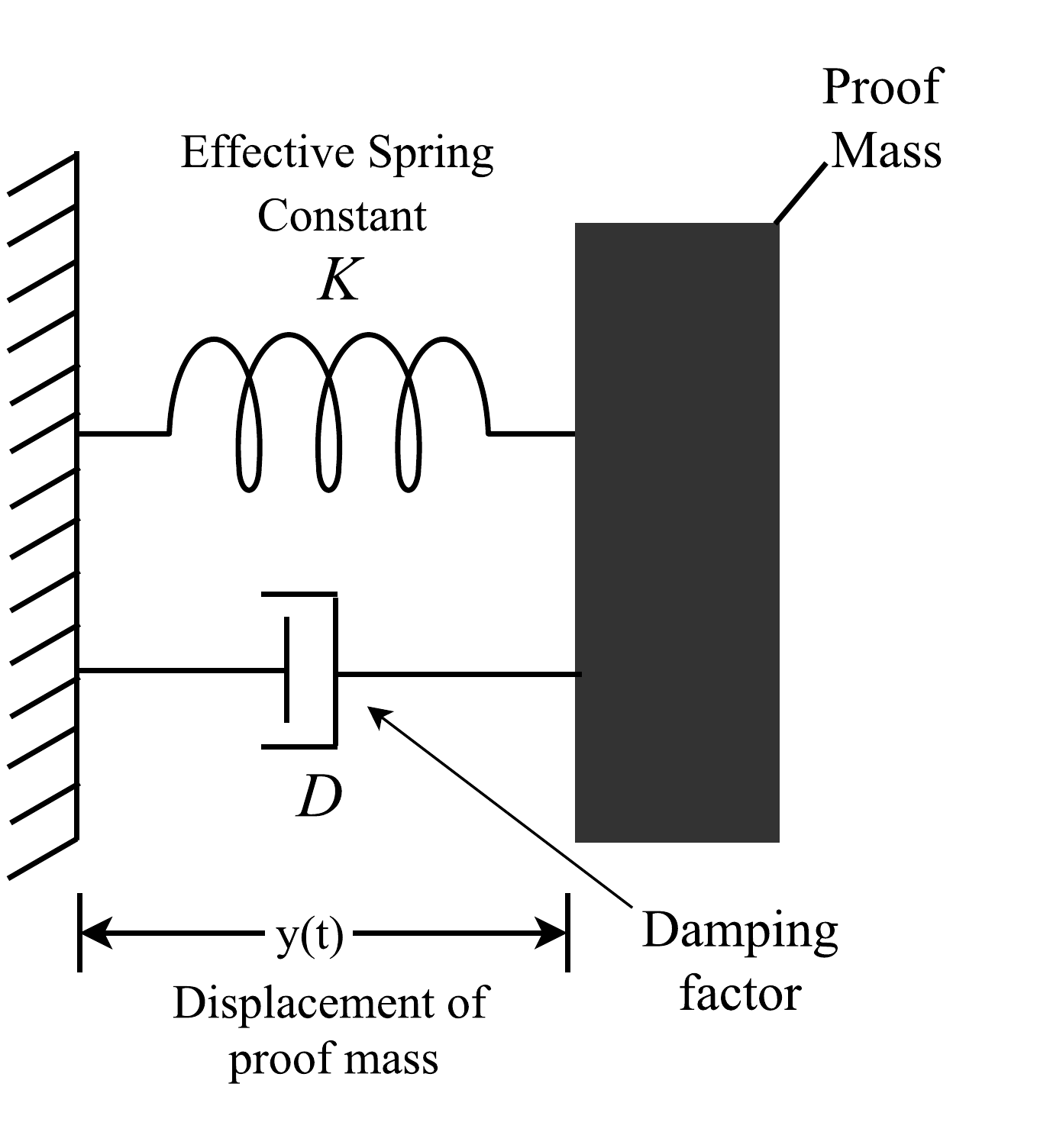}
  \caption{\hspace*{2.5cm}(b)}
  \label{FIG:mems_accelerometer}
\end{subfigure}
\caption{MEMs Accelerometer (a) Design (b) Mechanical Lumped Model}
\label{FIG:mems_accelerometer_total}
\end{figure}

An accelerometer is an electromechanical device that is used for the measurement of the static and the dynamic accelerations, i.e., the acceleration due to gravity anywhere on the earth and the acceleration due to the motion or vibration of an object according to the theory of relatively. It uses the sensors, which further make use of the environmental physical parameters, i.e., pressure, temperature, light and force.
Micro-Electro-Mechanical systems (MEMs) based accelerometer~\citep{kaajakari2009practical} are widely used as accelerometers. They are smaller in size, utilizing low power and thus due to these features, are integrated in a variety of applications, such as aircrafts~\citep{kuznetsov2011development}, airbag deployment~\citep{galvin2001microelectromechanical}, robotic telepresence~\citep{hung2004telepresence}, handheld computing gadgets~\citep{fennelly2012thermal}, natural disaster measurement devices~\citep{hsieh2014low} and automated external defibrillators (AEDs)~\citep{eggers2016wearable}.
Due to their wide usage in the safety critical domains, the accuracy of their frequency response analysis is of utmost importance. A typical MEMs accelerometer is depicted in
Fig.~\ref{FIG:mems_accelerometer_diagram}
whereas its mechanical lumped model~\citep{haykin2007signals} is illustrated in Fig.~\ref{FIG:mems_accelerometer}.

The differential equation modeling the dynamical behaviour of the MEMs accelerometer can be expressed as~\citep{haykin2007signals}:

\begin{equation}\label{EQ:diff_eq_mems_accelerometer}
 \dfrac{d^2y(t)}{dt^2} + \dfrac{D}{M}\dfrac{dy(t)}{dt} + \dfrac{K}{M} y(t) = u(t),
\end{equation}

In the above equation, $M$ is the proof mass, whereas, $K$ is the effective spring constant and $D$ represents the damping factor, which affects the dynamic movement of the proof mass as shown in Figure~\ref{FIG:mems_accelerometer}. All of these are design parameters of the underlying system and can have positive values only. Similarly, $x(t)$ is the external acceleration due to motion of the proof mass, whereas $y(t)$ is the displacement of the corresponding mass.

The corresponding frequency response of the MEMs accelerometer is given as follows:

\begin{equation}\label{EQ:freq_res_mems_accelerometer}
 \frac{Y(\omega)}{U (\omega)} = \dfrac{1}{{(i\omega})^2 + \dfrac{D}{M}(i\omega) + \dfrac{K}{M}}
\end{equation}

In order to verify its frequency response, we first model the corresponding differential equation as the following HOL-Light function:

\begin{defn}
\label{DEF:diff_equation_mems_accelerometer}
\emph{Differential Equation of MEMs Accelerometer} \\{\small
\textup{\texttt{$\vdash$  $\forall$ K D M. outlst\_de\_ma K D M = [K / M; D / M; \&1]
}}} \\
{\small
\textup{\texttt{$\vdash$ inlst\_de\_ma = [\&1]
}}} \\
{\small
\textup{\texttt{$\vdash$ diff\_eq\_MEMs\_ACC inlst\_de\_ma outlst\_de\_ma u y t K D M $\Leftrightarrow$ \\
$\mathtt{\ }$\hspace{2.25cm}  differential\_equation 2 (outlst\_de\_ma K D M) y t = \\
$\mathtt{\ }$\hspace{2.25cm} differential\_equation 0 inlst\_de\_ma u t
}}}
\end{defn}

\noindent where the function $\texttt{diff\_eq\_MEMs\_ACC}$ accepts the function variables $\texttt{x}$ and $\texttt{y}$ and the lists of coefficients ($\texttt{inlst\_de\_ma}$ and $\texttt{outlst\_de\_ma}$) and returns the corresponding differential equation of the MEMs accelerometer.

Next, we verify its frequency response as the following theorem in HOL-Light:

\begin{flushleft}
\begin{thm}
\label{THM:freq_response_mems_accelerometer}
\emph{Frequency Response of MEMs Accelerometer} \\{\small
\textup{\texttt{$\vdash$ $\forall$ y u w K D M.  \&0 < M $\wedge$  \&0 < D $\wedge$  \&0 < K $\wedge$\\
$\mathtt{\ }$\hspace{0.0cm}  ($\forall$t. differentiable\_higher\_derivative 2 y t) $\wedge$ ($\forall$t. u differentiable at t) $\wedge$  \\
$\mathtt{\ }$\hspace{0.0cm}  fourier\_exists\_higher\_deriv 2 y $\wedge$ fourier\_exists u $\wedge$  \\
$\mathtt{\ }$\hspace{0.0cm} ($\forall$k. k < 2 $\Rightarrow$  \\
$\mathtt{\ }$\hspace{1.0cm}(($\lambda$t. higher\_vector\_derivative k y $\mathtt{\overline{t}}$) $\rightarrow$ vec 0) at\_posinfinity) $\wedge$ \\
$\mathtt{\ }$\hspace{0.0cm}  ($\forall$k. k < 2 $\Rightarrow$  \\
$\mathtt{\ }$\hspace{1.0cm}(($\lambda$t. higher\_vector\_derivative k y $\mathtt{\overline{t}}$) $\rightarrow$ vec 0) at\_neginfinity) $\wedge$ \\
%$\mathtt{\ }$\hspace{0.0cm} (($\lambda$t. u (lift t)) $\rightarrow$ vec 0) at\_posinfinity $\wedge$ \\
%$\mathtt{\ }$\hspace{0.0cm} (($\lambda$t. u (lift t)) $\rightarrow$ vec 0) at\_neginfinity $\wedge$ \\
$\mathtt{\ }$\hspace{0.0cm}  ($\forall$t. diff\_eq\_MEMs\_ACC inlst\_de\_ma outlst\_de\_ma u y t K D M) $\wedge$  \\
$\mathtt{\ }$\hspace{0.0cm}  $\sim$(fourier u w = Cx (\&0)) $\wedge$ \\
$\mathtt{\ }$\hspace{0.0cm}  $\sim$((ii $\ast$ Cx w) pow 2 + Cx (D / M) $\ast$ ii $\ast$ Cx w + Cx (K / M) = Cx (\&0))  \\
$\mathtt{\ }$\hspace{0.2cm} $\Rightarrow$ fourier y w / fourier u w =  \\
$\mathtt{\ }$\hspace{1.5cm} Cx (\&1) / ((ii $\ast$ Cx w) pow 2 + Cx (D / M) $\ast$ ii $\ast$ Cx w + Cx (K / M))
}}}
\end{thm}
\end{flushleft}

The first three assumptions ensure that the variables corresponding to proof mass ($\texttt{M}$), spring constant ($\texttt{K}$) and damping factor ($\texttt{D}$) cannot be negative or zero. The next assumption ensures that the function $\texttt{y}$ is differentiable up to the second order. Similarly, the next assumption represents the differentiability condition for the function $\texttt{u}$. The next assumption represents the Fourier transform existence condition up to the second order derivatives of the function $\texttt{y}$. Similarly, the next assumption provides the Fourier transform existence condition of the function $\texttt{u}$. The next two assumptions represent the condition $ \lim\limits_{t \to \pm\infty} {y^{(k)} (t)} = 0 $ for each $ k = 0, 1 $, i.e., $ \lim\limits_{t \to \pm\infty}  {y^{(1)} (t)} = 0 $ and $ \lim\limits_{t \to \pm\infty}  {y^{(0)} (t)} = \lim\limits_{t \to \pm\infty}  y(t) = 0 $, where $y^{(k)}$ is the $k^{th}$ derivative of $\texttt{y}$.
The next assumption represents the formalization of Equation \ref{EQ:diff_eq_mems_accelerometer} and the last two assumptions provide some interesting design related relationships, which must hold for constructing a reliable MEMs accelerometer. Finally, the conclusion of the above theorem represents the frequency response, given by Equation \ref{EQ:freq_res_mems_accelerometer}. The proof is based on Theorem~\ref{THM:freq_response_nth_order_lti_system} along with some arithmetic reasoning.

Besides the above-mentioned audio equalizer and a MEMs based accelerometer applications, we also used the proposed formalization to formally verify the frequency response of a drug therapy model, which can be useful towards finding out a particular amount of dosage of drug lidocaine that has to be supplied to a particular person having the Ventricular arrhythmia and the details about its verification can be found in proof script~\citep{adnan16contsystemfouriertrans}.

\section{Discussion} \label{SEC:discussion}

The distinguishing feature of our proposed formalization as compared to traditional analysis methods is that all of the theorems verified are of generic nature, i.e., all of the variables and functions are universally quantified and can thus be specialized in order to obtain the results for some given values. Moreover, all of the required assumptions are guaranteed to be explicitly mentioned along with the theorem due to the inherent soundness of the theorem proving approach. Similarly, the verification of the frequency response described in Section~\ref{SEC:formal_analysis_generic_n_order_sys}, is of a generic $n$-order system, can be specialized in order to formally analyse any real-world system as presented in Section~\ref{SEC:applications}. Whereas, in the computer simulation techniques, we have to model each of the system individually.
Moreover, the high expressiveness of the higher-order logic enables us to model the differential equation and the corresponding frequency response in their true continuous form, whereas, in model checking they are mostly discretized and modeled using a state-transition system, which may compromise the accuracy of the analysis.

The above-mentioned formalization is done interactively.
However, we tried to automate the proof processes by making some simplification tactics. We develop a tactic \texttt{ASM\_REAL\_SIMP\_TAC}, which simplifies an expression involving real arithmetics using all of the assumptions of the theorem. We also develop a tactic \texttt{ASM\_COMPLEX\_SIMP\_TAC}, which simplifies a complex expression involving arithmetic operations using all of the assumptions of a theorem. We developed some more simplification tactics that can be found in our proof script~\citep{adnan16contsystemfouriertrans}.

The major difficulty faced during the formalization was the unavailability of detailed proofs for the properties of Fourier transform in literature. The available paper-and-pencil based proofs were found to be very abstract and missing the complete reasoning about the steps. The other challenge in reported formalization was that some of the assumptions of the properties of the Fourier transform were not mentioned in the literature as in the case of the first-order and higher-order differentiation properties, the assumptions $4$ and $5$ of theorem of \textit{First-order Derivative}, presented in Table~\ref{TAB:properties_of_Fourier_transform}, and the assumptions $3$ and $4$ of the \textit{Higher-order Derivative}, presented in Table~\ref{TAB:properties_of_Fourier_transform}, were absent from most of the analysis books.
The effort involved in the verification of individual theorem in the form of proof lines and the man-hours is presented in Table~\ref{TAB:verification_details_each_thm}.

%---------------------------------------------------------------------------------------%
%---------------------------------------------------------------------------------------%

\begin{table}[h]
\centering
\captionsetup{justification=centering}
\caption{Verification Detail for Each Theorem}
\label{TAB:verification_details_each_thm}
  \resizebox{1.0\textwidth}{!}{\begin{minipage}{\textwidth}
{\renewcommand{\arraystretch}{1.005}% for the vertical padding
\begin{tabular}{p{10.4cm} p{0.8cm} p{0.8cm}}
\hline\hline
 Formalized Theorems & Proof Lines  & Man-hours  \\ \hline \hline
  Theorem~\ref{THM:prop_01_integrable_univ} (Integrability of Improper Integral) & 880 & 87 \\ \hline
    Table~\ref{TAB:properties_of_Fourier_transform}: Linearity & 115 & 13 \\ \hline
  Table~\ref{TAB:properties_of_Fourier_transform}: Time Shifting & 190 & 22 \\ \hline
    Table~\ref{TAB:properties_of_Fourier_transform}: Frequency Shifting & 24 & 4 \\ \hline
  Table~\ref{TAB:properties_of_Fourier_transform}: Modulation & 98 & 11 \\ \hline
    Table~\ref{TAB:properties_of_Fourier_transform}: Time Scaling and Time Reversal & 200 & 25 \\ \hline
%  Table~\ref{TAB:properties_of_Fourier_transform}: Time Reversal & 78 & 17 \\ \hline
    Table~\ref{TAB:properties_of_Fourier_transform}: First Order Differentiation in Time Domain & 355 & 29 \\ \hline
  Table~\ref{TAB:properties_of_Fourier_transform}: Higher Order Differentiation in Time Domain & 110 & 15 \\ \hline
    Table~\ref{TAB:properties_of_Fourier_transform}: Area under a function & 12 & 1 \\ \hline
Theorems~\ref{THM:fourier_even_function} and~\ref{THM:fourier_odd_function}  (Relationship with Fourier Cosine and Fourier Sine Transforms) & 296 & 35 \\ \hline
    Theorem~\ref{THM:relation_fourier_laplace}  (Relationship with Laplace Transform) & 162 & 26 \\ \hline
Theorem~\ref{THM:fourier_transform_of_diff_equation}  (Fourier Transform of Differential Equation of Order \textit{n}) & 170 & 25 \\ \hline
    Theorems~\ref{THM:fourier_rect_pulse},~\ref{THM:fourier_one_sided_neg_exp},~\ref{THM:fourier_bilateral_exp},~\ref{THM:fourier_sine_tone_burst} and~\ref{THM:damped_one_sided_sine} (Fourier Transform of Some Commonly Used Functions) & 275 & 35 \\ \hline
    Theorem~\ref{THM:freq_response_nth_order_lti_system} (Frequency Response of n-order LTI System) & 65 & 12 \\ \hline
        Theorem~\ref{THM:freq_response_audio_equalizer_3} (Frequency Response of an Audio Equalizer) & 90 & 10 \\ \hline
            Theorem~\ref{THM:freq_response_mems_accelerometer} (Frequency Response of MEMs Accelerator) & 63 & 6 \\ \hline
\end{tabular}
  }
%   }
      \end{minipage}}
\end{table}

%---------------------------------------------------------------------------------------%
%---------------------------------------------------------------------------------------%

\noindent The proof process for the formal verification of Theorems~\ref{THM:freq_response_nth_order_lti_system},~\ref{THM:freq_response_audio_equalizer_3} and~\ref{THM:freq_response_mems_accelerometer} took only 218 lines and 28 man-hours and was very simple and straightforward compared to the reasoning process of Theorem~\ref{THM:prop_01_integrable_univ}, \textit{first-order differentiation}, \textit{higher-order differentiation} properties presented in Table~\ref{TAB:properties_of_Fourier_transform} and Theorem~\ref{THM:fourier_transform_of_diff_equation}, which involves more effort and user interaction. This clearly illustrates the benefits of our foundational formalization, presented in Section~\ref{SEC:Formal_verif_Fourier_properties} of this paper. Moreover, the man-hours are calculated based on two factors. The first factor includes the number of lines of code per hour by a person with an average expertise and the second factor is the difficulty of the proof. For example, the proof lines for Theorem~\ref{THM:freq_response_nth_order_lti_system} and~\ref{THM:freq_response_mems_accelerometer} are almost same, whereas the man-hours for both of the theorems are different, i.e., the man-hours for Theorem~\ref{THM:freq_response_nth_order_lti_system} are double in number with respect to man-hours for Theorem~\ref{THM:freq_response_mems_accelerometer}.

\section{Conclusions}\label{SEC:Conclusion}

In this paper, we proposed a formalization of Fourier transform in higher-order logic in order to perform the frequency domain analysis of the continuous-time systems. We presented the formal definition of Fourier transform and based on it, verified its classical properties, namely existence, linearity, time shifting, frequency shifting, modulation, time reversal, time scaling, differentiation and its relationship to Fourier Cosine, Fourier Sine and Laplace transforms. We also presented the formal verification of some commonly used functions. Next, we provided the formal verification of the frequency response of a generic \textit{n}-order system. Lastly, in order to demonstrate the practical effectiveness of the proposed formalization, we presented a formal analysis of an audio equalizer and a MEMs accelerometer.

In the future, we aim to verify the two-dimensional Fourier transform~\citep{bracewell1965fourier}, which is frequently applied for the frequency-domain analysis of many optical systems, electromagnetic theory and image processing algorithms. We also plan to formalize the inverse Fourier transform and verification of its properties, which would be very helpful to reason about the solutions to differential equations. This formalization can be further used in our project on system biology~\citep{arashid2017sbiology}, to
formally analyze the differential equations corresponding to the reaction kinetic models of the biological systems.

\section*{Acknowledgements}

This work was supported by the National Research Program for Universities grant (number 1543) of Higher Education Commission (HEC), Pakistan.

\bibliographystyle{elsart-harv}
\bibliography{bibliotex}

\end{document}